\documentclass[aps,pra,twocolumn,amsmath,amssymb,superscriptaddress,reprint,longbibliography]{revtex4-1}

\usepackage[utf8]{inputenc}
\usepackage{graphicx}
\usepackage{color}
\usepackage[colorlinks=true,citecolor=blue]{hyperref}
\usepackage{units}
\usepackage{amsmath}
\usepackage{tabularx}
\usepackage{lipsum}
\usepackage{float}

\usepackage{amssymb}
\usepackage{xcolor}
\usepackage{extarrows}
\usepackage{tikz-cd}

\newcommand{\bsigma}{\boldsymbol{\sigma}}
\newcommand{\bvarsigma}{\boldsymbol{\varsigma}}

\AtBeginDocument{%
 \newwrite\bibnotes
 \def\bibnotesext{Notes.bib}
 \immediate\openout\bibnotes=\jobname\bibnotesext
 \immediate\write\bibnotes{@CONTROL{REVTEX41Control}}
 \immediate\write\bibnotes{@CONTROL{%
 apsrev41Control,author="08",editor="1",pages="1",title="0",year="1"}}
 \if@filesw
 \immediate\write\@auxout{\string\citation{apsrev41Control}}%
 \fi
}%

\begin{document}

\title{Impact of Interparticle Interaction on Thermodynamics of Nano-Channel Transport of Two Species}
\author{Wolfgang R. Bauer}
\affiliation{ \quad Department of Internal Medicine I, University Hospital of W\"urzburg, Oberd\"urrbacher Stra{\ss}e 6, D-97080 W\"urzburg, Germany; bauer\_w@ukw.de\\
\quad Comprehensive Heart Failure Centre, Am Schwarzenberg 15, A15, D-97080 W\"urzburg, Germany}

\date{\today}

\begin{abstract}

Understanding the function and control of channel transport is of paramount importance for cell physiology and nanotechnology. In particular, if several species are involved, the mechanisms of selectivity, competition, cooperation, pumping, and its modulation need to be understood. What lacks is a rigorous mathematical approach within the framework of stochastic thermodynamics, which explains the impact of interparticle in-channel interactions on the transport properties of the respective species. To achieve this, stochastic channel transport of two species is considered in a model, which different from mean field approaches, explicitly conserves the spatial correlation of the species within the channel by analysis of the stochastic dynamics within a state space, the elements of which are the channel's spatial occupation states. The interparticle interactions determine the stochastic transitions between these states. Local flow and entropy production in this state space reveal the respective particle flows through the channel and the intensity of the Brownian ratchet like rectifying forces, which these species exert mutually on each other, together with its thermodynamic effectiveness and costs. Perfect coupling of transport of the two species is realized by an attractive empty channel and strong repulsive forces between particles of the same species. This confines the state space to a subspace with circular topology, in which the concentration gradients as thermodynamic driving forces act in series, and channel flow of both species becomes equivalent. For opposing concentration gradients, this makes the species with the stronger gradient the driving, positive entropy producing one; the other is driven and produces negative entropy. Gradients equal in magnitude make all flows vanish, and thermodynamic equilibrium occurs. A differential interparticle interaction with less repulsive forces within particles of one species but maintenance of this interaction for the other species adds a bypass path to this circular subspace. On this path, which is not involved in coupling of the two species, a leak flow of the species with less repulsive interparticle interaction emerges, which is directed parallel to its concentration gradient and, hence, produces positive entropy here. Different from the situation with perfect coupling, appropriate strong opposing concentration gradients may simultaneously parallelize the flow of their respective species, which makes each species produce positive entropy. The rectifying potential of the species with the bypass option is diminished. This implies the existence of a gradient of the other species, above which its flow and gradient are parallel for any gradient of the less coupled species. The opposite holds for the less coupled species. Its flow may always be rectified and turned anti-parallel to its gradient by a sufficiently strong opposing gradient of the other one.

\end{abstract}

\maketitle
\begin{widetext}

\section{Introduction}
Channel transport of particles connecting otherwise separated environments is of paramount importance for regulation of cellular, sub-cellular, and molecular processes but also an emerging field of research in nanotechnology. According to this importance, there exists an abundance of work addressing how the effectiveness and selectivity of this channel transport may be modulated and increased. Much this work focuses on models that exploit very detailed information about channel structure and channel particle interaction to answer questions about real channels. Others, as we, have more fundamental aspects in mind to understand the basic thermodynamic properties of the channel. 

Of the latter, many manuscripts addressed the impact of the particle channel interaction on the effectiveness and selectivity of transport. Focus laid on static \cite{Berezhkovskii2005, Berezhkovskii2005b, bauer2005, Bauer2006, Kolomeisky} but also on temporally modulated interactions with the channel, e.g., if stochastic gating plays a role \cite{Berezhkovskii_2018, Carusela_2018, davtyan2019theoretical,lisowski2019entropy}. In contrast, the role of the interparticle interactions on channel transport, especially for the case that several species are involved, leaves many open questions. Simulations \cite{Zilman2007PLoSComputBiol, Zilman2010PloSComputBiol} demonstrated a potential cooperation of two species within the channel, but the mechanism behind was not revealed. One-dimensional exclusion models of two species channel transport showed that with increasing channel length, osmosis and related processes that rely on interparticle interactions become more effective \cite{Chou1, Chou2, Chou3, Bauer2013, Bauer2017}. Mean field approximations addressed how jamming of a single species inside the channel affects the transport parameters; however, though qualitatively correct, results differed from simulations for narrow channels \cite{Zilman2009PRL, Zilman2010}. The clear drawback of mean field theories is that they derive a mean interparticle interaction from a mean occupation probability of particles, i.e., spatial correlations between particles are neglected. However, an interparticle interaction definitely implies a strong correlation between occupation states within its spatial range, which makes mean field theories only applicable for very short range interactions.  

What is still left is a rigorous, mathematical approach that is exactly solvable and addresses the effect of interparticle interaction on transport without the method inherent constraint of mean field theories in terms of stochastic thermodynamics. A model within this framework is a prerequisite for understanding the channel transport of two species and their mutual effect on each other beyond just a phenomenological descriptive approach, as provided by simulations. This is the aim of this manuscript. As the interparticle interaction is addressed, spatial correlations between particles in the channel must be conserved. This is achieved by mapping the dynamics of particle transport on the transition dynamics of occupation states in the channel, which form the state space. The probability of these states then directly reflects how particles are spatially correlated. Analysis of transition dynamics in this state space will allow in a unique way seeing how the effects of the driving forces of channel transport, namely the particle concentrations in the baths adjacent to the channel ends, are distributed within this space. This will elucidate the mechanisms by which the driving force of one species affects the transport of the other and vice versa. Furthermore, the thermodynamic sources of this complex channel transport, i.e., regions of positive entropy production, may be allocated in state space. It becomes obvious how transitions in state space and the respective probability flows are related to these sources of entropy production and how interparticle interactions direct these sources to achieve mutual rectifying forces, which in the case of opposing concentration gradients, makes entropy sinks, i.e., regions of negative entropy production, emerge.  

In this sense, the manuscript is organized as follows. In Section \ref{sec2}, we present the mathematical framework. A brief presentation of the channel model and state space is followed by the description of the ratchet mechanism by which particles mutually exert rectifying forces on each other and how this translates into stochastic thermodynamics. It is analyzed how the thermodynamic forces drive the system within the network of state space and how the local parameters of state space as flow between states and associated entropy sources are related to particle flow and global entropy production. With these tools, we analyze in Section \ref{sec3} how modulation of intra-species interparticle interactions confines state space by optimal coupling of transport, which achieves a maximum rectification. In Section \ref{sec4}, the constraint of strict coupling is lowered for one species, which expands the confined state space. The consequences for the rectification capability become evident in phase diagrams, in which for each species, the parallel and anti-parallel direction of concentration gradients and particle flows define different phases, whose transitions depend on the concentration gradients of both species.

\section{The Model and Stochastic Thermodynamics in State Space} \label{sec2}
\subsection{State Space and Transition Rates Within}

A channel connects two baths, labeled $1$ on the right and $ 2$ on the left site (Figure~\ref{StateSpaces}). Each of them contains particles of species $A$ and $B$ with respective concentrations $c_{1/2}^{(A/B)}$. The baths are supposed to be on the same energy level, i.e., the only thermodynamic driving forces the particles are subjected to are the concentration gradients. The system is perfectly thermostat controlled at a temperature $T$, which allows normalizing all energetic quantities to $kT$ and leaves them dimensionless. The channel is narrow in the sense that only one particle can stay at a particular position along the channel axis. There exist only discrete channel positions along the channel axis, which are numbered $N\hdots2,\;1$. The occupation state of a channel is then described by a state variable $\boldsymbol{\sigma}=(\sigma_N\cdots,\sigma_2,\; \sigma_1) $, where $\sigma_i$ may take the values $\sigma_i=A,\;B,\; 0$ depending on whether position $i$ is occupied by species $A$, $B$, or non-occupied ($=0$), respectively. These states form the state space $\boldsymbol{\Sigma}=\{\boldsymbol{\sigma}\}$. This implies for a channel with a length of $N$ positions a state space with $3^N$ elements. In this manuscript, we restrict the channel lengths to $N=2$ or 3 positions, i.e., the state space has either 9 ($N=2$) or 27 elements ($N=3$), as shown in Figure~\ref{StateSpaces}. 
Transitions dynamics results from a superposition of interaction forces (interparticle and particle-channel) and stochastic forces. The latter determine the random access of particles from the baths to free channel ends or random jumps within the channel to nearest free neighbor positions. This dynamics is described by a stationary Markov process, i.e., the evolution of the probability $P_{\boldsymbol{\sigma}}(t)$ to find the channel in the state $\boldsymbol{\sigma}$ at time $t$ obeys the master equation \cite{Gardiner}, 
\begin{equation}
\dot{P}_{\boldsymbol{\sigma}}(t)=\sum_{\boldsymbol{\varsigma}\in \boldsymbol{\Sigma}} \lambda_{\boldsymbol{\sigma},\boldsymbol{\varsigma}} P_{\boldsymbol{\varsigma}}\;,\label{Master}
\end{equation} 
with transition rates $\lambda_{\boldsymbol{\sigma},\boldsymbol{\varsigma}}=\lambda_{\boldsymbol{\sigma}\leftarrow\boldsymbol{\varsigma}}$ from state $\boldsymbol{\varsigma}$ to state $\boldsymbol{\sigma}$, which comprise the above-mentioned forces. Conservation of probability, $\sum_{\bsigma} P_{\bsigma}=1$, yields the diagonal matrix elements as $\lambda_{\bvarsigma,\bvarsigma}=-\sum_{\bsigma\neq\bvarsigma} \lambda_{\bsigma,\bvarsigma}$. This enables one to rewrite the master equation in the form or a continuity equation: 
\begin{equation}
\dot{P}_{\bsigma}=\sum_{\bvarsigma} J_{\bsigma,\bvarsigma}\;,\label{Continuity}
\end{equation} 
with: 
\begin{equation}
J_{\bsigma, \bvarsigma}=\lambda_{\boldsymbol{\sigma},\boldsymbol{\varsigma}} P_{\boldsymbol{\varsigma}}-\lambda_{\boldsymbol{\varsigma},\boldsymbol{\sigma}} P_{\boldsymbol{\sigma}}\label{LocalFlow}
\end{equation}
as flows of probability in state space from state $\bvarsigma$ to state $\bsigma$. Note that by taking the occupation states $\boldsymbol\sigma$ of the channel as the state variable, spatial correlations related to interparticle interaction become explicit in the probability $P_{\boldsymbol\sigma}$. This is in contrast to mean field theories, which consider a mean occupation probability of a species at some position in the channel as the base to define a mean interaction force. 
\begin{figure}[H]
\centering
\includegraphics[width=15cm]{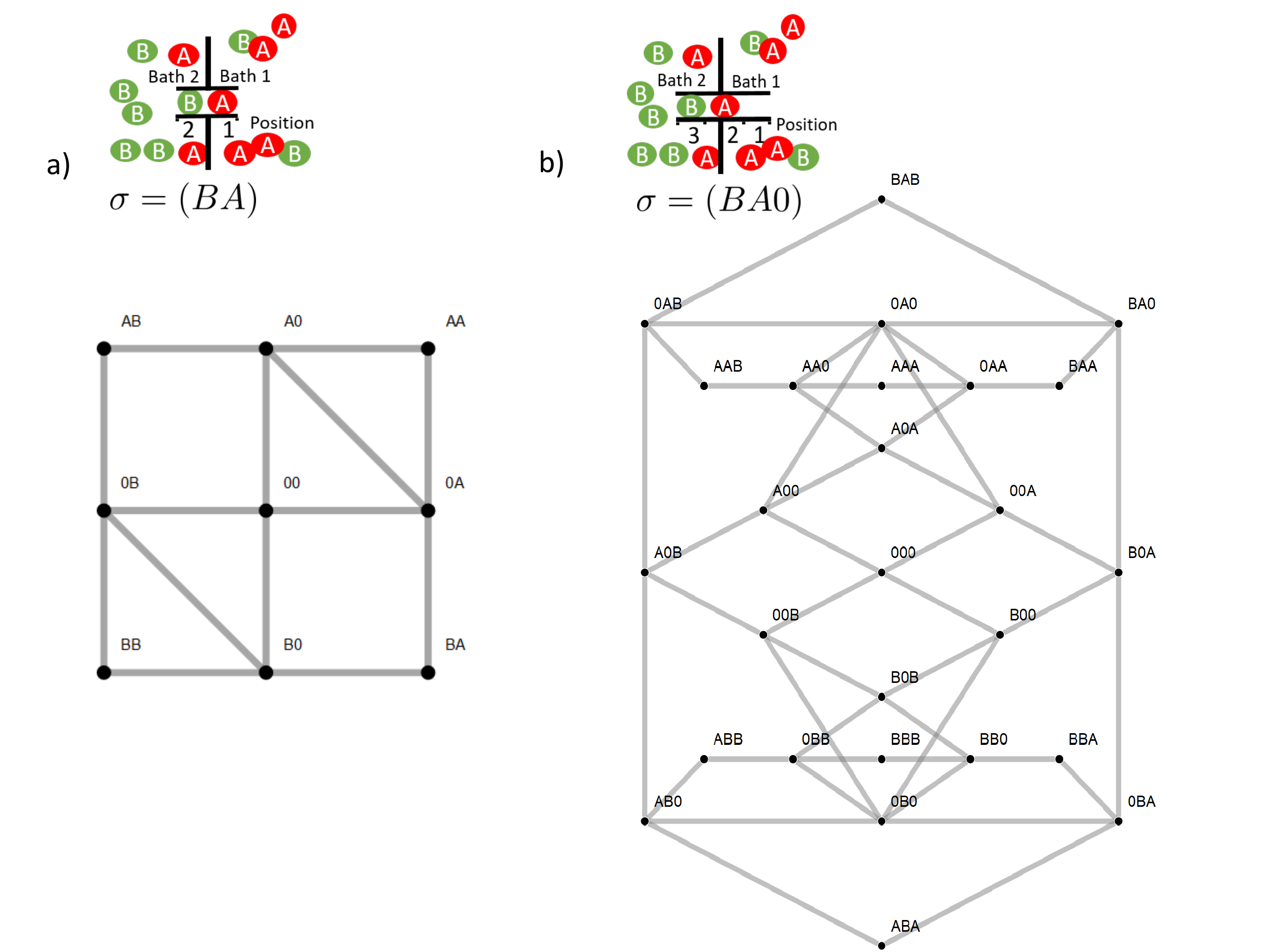}
\caption{Sketch of a two site ({\bf a}) and three site channel ({\bf b}) with example state $\bsigma$ and the state space below. The two site space has nine and the three site space 27 elements (see text). States between which exist stochastic transitions are linked by a gray line. The respective rates are given in the text. } 
\label{StateSpaces}
\end{figure}

To obtain the transition rates $\lambda$, we have to differentiate between particle-channel and interparticle interactions. We assume no differential particle-channel interaction forces inside the channel, or in other words, the corresponding energy profile is flat. Hence, a single particle inside an otherwise empty channel may hop with the same rate $\tau_0^{-1}$ to its nearest neighbor positions, with $\tau_0$ as the corresponding time constant. In this manuscript, all dynamical processes will be normalized to this baseline dynamics, i.e., all rates $\lambda_ {\bsigma,\bvarsigma}$ are given as dimensionless multiples of $\tau_0^{-1}$. Access of particles of some species being in some bath with concentration $c$ to a free channel end position shall be given by a baseline rate $k_+ c$ and conversely hopping away from the channel end into the bath by the rate $k_-$. These rates are identical for both species and baths. As we are mainly interested in interparticle interactions, we assume that the probability to find an empty channel is very low. In other words, in addition to the baseline in and out hopping rates, the empty channel has a very strong affinity to particles of either species. Formally, this is achieved by assigning absorption of particles by the empty channel, i.e., transitions from $\bsigma=(00)$ for the two site or $\bsigma=(000)$ for the three site channel, an energy gain $-E_{0}<0$. Repulsive interparticle interactions result from the constraint that an occupied position may not be occupied further, neither from the baths, nor from inside the channel. However, as we are interested also in more subtle interparticle interactions, we further introduce repulsive forces between particles of the same species inside the channel. These repulsive forces hamper the access of particles from the baths to the channel if the channel is already occupied by the same species. Oppositely, these forces make it easier for a particle to leave the channel, if the channel is already occupied by particles of the same species. For the two site channel model, this is simply achieved by assigning channel states $\bsigma=(XX)$, which are occupied by two particles of the same species $X$, a higher energy level $E_{X}>0$. For the longer three site channel, we differentiate between a short- (sr) and a long-range (lr) repulsive interaction. The long-range interaction is similar to the situation of the two site channel, i.e., once a particle is inside the channel, it, independently of its position, rejects access of particles of the same species. In contrast, the short-range interaction reveals a spatial dependence. It rejects the access of particles of the same species, which potentially could become its nearest neighbor, e.g., $(0X0)\to(XX0)$, but there is no repulsive force when there is a vacant position in between, e.g., $(00X)\to(X0X)$. It is obvious that these interparticle interactions facilitate the occupation of the channel by particles of different species and, by this favor, the option of cooperation or competition between them. 
Formally, these interactions modulate the jump in rate into the empty channel  and the corresponding jump out rate by \cite{Nadler86}:
\begin{eqnarray}
k_+ c &\to& e^{E_0/2}\; k_+ c \cr
k_-&\to& e^{-E_0/2}\; k_- \;, \label{E0}
\end{eqnarray} 
and similarly, a particle's access to and departure from a channel already occupied by the same species by:
\begin{eqnarray}
k_+ c^{(X)} &\to& e^{-E_X/2}\; k_+ c^{(X)} \cr
k_-&\to& e^{E_x/2} \; k_-\;, \label{EX}
\end{eqnarray} 
if the interaction is present. In the case of the three site channel, this always holds for the long-range and potentially (see above) for the short-range interaction. Note that these modulations of rates fulfill the detailed balance condition, i.e., for the ratio of jump in and out rates, $\sim e^{-\delta E}$ holds, with $\delta E=E_X\;\hbox{or}\; - E_0$ as the energy difference between the two channel states.

\subsection{The Ratchet Mechanism of Interspecies Interaction and Stochastic Trajectories in State Space}

As a position inside the channel may only be occupied by one particle, states with neighboring particles of different species may only undergo transition towards states in which a particle has moved oppositely to the position of its neighbor. In the case of vanishing concentration gradients, i.e., in thermodynamic equilibrium, the stochastic dynamics is symmetrical along the channel axis. In particular, this holds for the occupation of the channel with neighboring particles of different species. Hence, the above constraint of particle motion is balanced, i.e., as expected for thermodynamic equilibrium, these constraints do not induce the flow of particles. However, the situation changes in the presence of a concentration gradient of one species while maintaining a vanishing gradient for the other one. The probability to find a particle of the species with a non-vanishing gradient inside the channel declines in the direction of this gradient. This symmetry break implies that due to interparticle interactions, the species with the vanishing gradient has more options to move in the direction of this concentration gradient, i.e., a net flow occurs (Figure~\ref{Riemann}). This is the key feature of the Brownian ratchet paradigm \cite{Cubero2016}, in which a driving/fluctuating asymmetric potential rectifies the motion of a particle. The fluctuating asymmetric potential corresponds to the concentration gradient related asymmetric stochastic access from the baths, which rectifies the motion of the species with the vanishing concentration gradient. If both species have a concentration gradient, each species is at the same time ratchet for the other and also rectified by subjection to the ratchet function of the other. 
This naive explanation of the ratchet mechanisms, however, blanks out the fact that the ratchet itself is subject to thermodynamics. The ratchet mechanism does not work for free, but takes its toll based on the second law of thermodynamics, which requires a positive net entropy production. This interwoven network of mutual ratchet function and subjection to rectification, and its degree of effectiveness, is best resolved within the framework of state space and its stochastic transitions within. 
Transitions between two states $\bsigma \rightleftharpoons\varsigma$ in state space are driven by the free energy difference between both, which is obtained from the transitions rates by \cite{Gardiner}:
\begin{equation}
\Delta\epsilon_{\bsigma,\bvarsigma}=-\ln(\lambda_{\bsigma,\bvarsigma}/\lambda_{\bvarsigma,\bsigma})\;.\label{FreeEnergyLocal}
\end{equation} 
Depending on whether the states refer to different energetic levels or particle uptake/release, the transition implies a change of the entropy of the baths \cite{Seifert_2012}, which is related to its heat or particle exchange with the channel: 
\begin{equation}
\Delta s_{ \text{baths}\; \bsigma,\bvarsigma}=-\Delta\epsilon_{\bsigma,\bvarsigma}\;. \label{LocalEntropyBathTraj}
\end{equation}
Note that we normalized all energetic quantities to temperature, so that the temperature does not appear in the entropy. 
Some transitions comprise both, heat and particle exchange, e.g., if an occupied channel hampers access of particles of the same species from the bath by a repulsive interparticle interaction. Therefore, the free energy comprises both, as well, as shown by insertion of the rates of Equations~(\ref{EX}) into Equation~(\ref{FreeEnergyLocal}). For this transition, the entropy change takes the form: 
\begin{equation}
\Delta s_{ \text{baths}}=\underbrace{-E_X}_{\text{heat exchange}}+\underbrace{\ln\left(\frac{k_+ c^{(X)}}{k_-}\right)}_{\text{particle exchange}}\label{LocalEntropyBathTraj2}
\end{equation} 
 
For an ensemble of channels, the entropy production rate related to the transition $\bsigma \rightleftharpoons\varsigma$ is then determined by the probability flow between the corresponding states (Equation~(\ref{LocalFlow})) as \cite{Schnackenberg}:
\begin{equation}
\dot{S}_{ \text{baths}\; \bsigma,\bvarsigma}= - J_{\bsigma, \bvarsigma} \Delta\epsilon_{\bsigma,\bvarsigma}\;. \label{LocalEntropy}
\end{equation}
This relationship first postulated by Schnackenberg relies on a perfectly working bath, i.e., heat and particle concentrations are instantaneously equilibrated, which we also assume here. Otherwise, it provides a lower bound of entropy production \cite{Ziener_2015}.

The free energy differences $\Delta\epsilon_{\bsigma,\bvarsigma}$ may be considered as a field of drift forces, superimposed on random forces, which affect the stochastic path through the state space. The free energy difference along such a path $\gamma=(\bsigma_N,\bsigma_{N-1},\hdots\bsigma_{i+1},\bsigma_i\hdots\bsigma_1)$ at ordered time points $t_i$ is then the sum:
\begin{equation}
\Delta \mathcal{E}_{\gamma}=\sum_{i=1}^{N-1}\Delta\epsilon_{\bsigma_{i+1},\bsigma_i}\;,\label{FreeEnergyGain}
\end{equation} 
with an according change of the entropy in the baths. 

We will first analyze this force field under equilibrium conditions, $\epsilon^{(eq)}_{\bsigma,\bvarsigma}$, i.e., if particle concentrations of each species are equal in the connected baths. Equilibrium implies that detailed balance makes the probability flow between two states in Equation~(\ref{LocalFlow}) vanish: 
\begin{equation}
J^{(eq)}_{\bsigma,\bvarsigma}=\lambda_{\boldsymbol{\sigma},\boldsymbol{\varsigma}} P^{(eq)}_{\boldsymbol{\varsigma}}-\lambda_{\boldsymbol{\varsigma},\boldsymbol{\sigma}} P^{(eq)}_{\boldsymbol{\sigma}}=0\;,
\end{equation}
with $P^{(eq)}$ as the equilibrium occupation probability distribution. Defining the potential:
\begin{equation}
\phi_{\bsigma}=-\ln(P^{(eq)}_{\boldsymbol{\sigma}})\label{Potentialeq}
\end{equation}
 yields with Equation~(\ref{FreeEnergyLocal}):
 \begin{equation}
 \Delta\epsilon^{(eq)}_{\bsigma,\bvarsigma}=\phi_{\bsigma}-\phi_{\bvarsigma}\;.\label{PotentialDiff}
 \end{equation}
This implies that the free energy difference along a stochastic path is simply given by the difference of the potentials between its ends $ \Delta \mathcal{E}_{\gamma}^{(eq)}=\phi_N-\phi_1$, i.e., in particular, the free energy difference for closed paths vanishes. This makes $\Delta\epsilon^{(eq)}_{\bsigma,\bvarsigma} $ a conservative field and defines the free energy landscape above state space $\boldsymbol\Sigma$ by the function $\phi: \boldsymbol\Sigma\ni\bsigma\to\phi_{\bsigma}$. 

The situation is different in the presence of concentration gradients and particle transport through the channel. A non-vanishing net transport of particles through the channel requires that the system visits repetitively states involved in particle exchange with the bath. Therefore, the stochastic path in this finite state space may be built from closed paths, which contain state transitions with the baths. The second law of thermodynamic implies that the free energy declines on most of these paths, as otherwise, on average, there would be no positive entropy production and, hence, no net particle flow. In contrast to the equilibrium situation, the free energy differences $\Delta\epsilon_{\bsigma,\bvarsigma}$ now form a non-conservative field, which successively drives the stochastic path towards lower free energy levels, and by this produces positive entropy in the baths. Thus, the free energy landscape cannot be described anymore by a potential function; instead, it is similar to a Riemann surface with logarithmic branching points (Figure~\ref{Riemann}). Note that ``successive'' for the free energy decline is not meant in the sense of monotonous. Of course, on the single trajectory level, there is the option of transient negative entropy production, i.e., increase of free energy. However, eventually, the free energy of the trajectory declines at arbitrary low values, $\lim_{t\to \infty} \Delta \mathcal{E}_{\gamma(t)}\to -\infty$. Furthermore, on the ensemble level, negative local entropy production for transitions in state space as given by Equation~(\ref{LocalEntropy}) is naturally feasible, but overall entropy production of the ensemble average in state space must be positive.  

\begin{figure}[h]
\centering
\includegraphics[width=13cm]{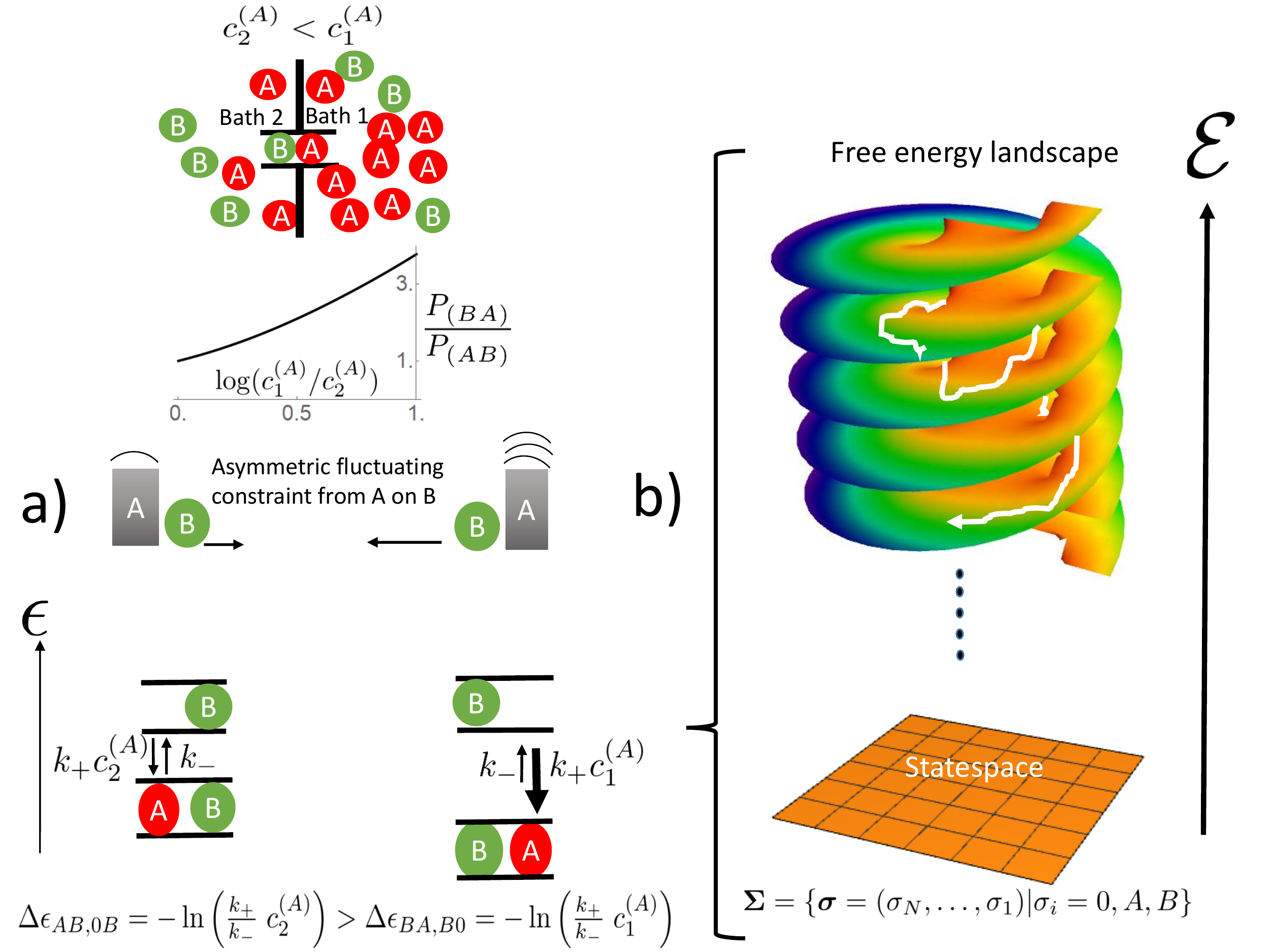}
\caption{The ratchet mechanism of state transitions ({\bf a}) and its integration into the free energy landscape on state space ({\bf b}). Left site (a) from above to below: A two site channel connecting two baths is shown with a concentration gradient of species $A$ directing from Bath 1 to Bath 2. The random access of species $A$ to the channel ends imposes fluctuating constraints on the mobility of species $B$. These are biased by the concentration gradient of $A$, which is evident from the ratio of state probabilities $P_{(BA)}/P_{(AB)}>1 $. Its dependence on the gradient of $A$ is shown in the graphic insert (probabilities were obtained from Equation~(\ref{Master}) for stationary conditions, $\dot{P}_{\bsigma}=0 $, and rates from Equations~(\ref{E0}) and (\ref{EX}), with $E_0=E_A=E_B=0$, $\;k_+c_1^{(B)}=k_+c_2^{(B)}=0.1$, $\;k_+c_2^{(A)}=0.1$, and $k_+c_1{(A)}$ varied). This bias favors movement of species $B$ in the direction of the gradient (longer arrow to the left). This biased fluctuating constraint of $A$ on $B$ translates into different free energy differences $\Delta\epsilon=-\ln(k_+\; c/k_- )$ of the states involved. Together with the affinity of the empty channel in Equation~(\ref{E0}) and the interaction energies of Equations~(\ref{EX}), they form the free energy landscape above the state space as shown in Sketch (b). For non-vanishing concentration gradients, this landscape cannot be obtained from a potential function as under equilibrium conditions from Equations~(\ref{Potentialeq}) and (\ref{PotentialDiff}). Instead it is similar to a Riemann surface with logarithmic branching points. This topology of the free energy landscape drives successively the stochastic trajectories (white) towards lower free energy levels $\mathcal{E}$. } 
\label{Riemann}
\end{figure}

\subsection{Local Probability Flows in State Space and Particle Flows through the Channel}
In an ensemble of channels, the dynamics of stochastic trajectories translates on average into probabilities of states and probability flows between these states (Equation~(\ref{LocalFlow})). 
In this manuscript, channel transport is studied in the steady (stationary) state, i.e., 
\begin{equation}
\dot{P}_{\bsigma}(t)=0 \label{steadystatesingle}
\end{equation}
holds. The corresponding probability distribution is then determined from Equation~(\ref{Master}) as:
\begin{equation}
\boldsymbol{\Lambda}\boldsymbol{P}^{(s)}=0\;, \label{steadystate}
\end{equation}
where we summarized the transition rates in the matrix $\boldsymbol{\Lambda}=(\lambda_{\bsigma,\bvarsigma})$ and the steady state probability in the vector $\boldsymbol{P}^{(s)}=(P^{(s)}_{\bsigma})$. Equation~(\ref{LocalFlow}) then determines the flow of probability between states. 

To obtain the particle flow through the channel from the probability flows in state space, one has to keep in mind that the 
steady state condition of Equation~ (\ref{steadystatesingle}) and the continuity Equation (\ref{Continuity}) imply the conservation of probability flows in state space:
\begin{equation}
\sum_{\bvarsigma} J^{(s)}_{\bsigma,\bvarsigma}=0\;.\label{Kirchhoff}
\end{equation}
This feature is well known from electrical circuits as Kirchhoff's law. To obtain the particle flow $J^{(X)}$ of species $X$ through the channel from flows in state space, it is in the steady state sufficient to consider state transitions, which are involved in particle exchange with the bath at some channel end. For example, for Bath 2 at the left site, these transitions are $\bsigma=(0,\hdots)\rightleftharpoons \bvarsigma=(X,\hdots)$. Flow then becomes:
\begin{equation}
J^{(X)}=\sum_{\bsigma, \varsigma| \hbox{\tiny{exchange $X$ with Bath 2}}} J^{(s)}_{\bsigma,\bvarsigma}\;.\label{FlowSpeciesv}
\end{equation} 
For the two site channels, an even simpler expression is obtained. States involved in the exchange of species $X$ with Bath 2 at the left site are $(X0),\; (XX),\; (XY)$, i.e., $J^{(X)}=J^{(s)}_{(00),(X0)}+J^{(s)}_{(0X),(XX)}+J^{(s)}_{(0Y),(XY)}$. A channel state with two sites occupied has only two options for transition, i.e., $(0X)\leftarrow(XX)\to (X0), \text{and}\; (0Y)\leftarrow (XY)\to(X0) $. Therefore, application of Kirchhoff's law implies $J^{(s)}_{(0X),(XX)}=-J^{(s)}_{(X0),(XX)}=J^{(s)}_{(XX),(X0)}$, and $J^{(s)}_{(0Y),(XY)}=-J^{(s)}_{(X0),(XY)}=J^{(s)}_{(XY),(X0)}$, where we exploited that the sign of flow changes concordant with the view of the direction of state transition, $J_{\bsigma,\bvarsigma}=-J_{\bvarsigma,\bsigma}$. Hence, $J^{(X)}=J^{(s)}_{(00),(X0)}+J^{(s)}_{(XX),(X0)}+J^{(s)}_{(XY),(X0)}$. Again, application of Kirchhoff's law, $J^{(s)}_{(00),(X0)}+J^{(s)}_{(XX),(X0)}+J^{(s)}_{(XY),(X0)}+J_{(0X),(X0)}=0 $, and respecting the change of sign when changing the transition direction yield:
\begin{equation}
J^{(X)}=J^{(s)}_{(X0),(0X)}\;.\label{FlowSpecies}
\end{equation}
Equations~(\ref{FlowSpeciesv}) and (\ref{FlowSpecies}) allow now the determination of particle flows through the channel.

\subsection{Sources of Entropy Production in State Space and Entropy Production by Channel Transport}

The fact that the driving forces, namely the concentration gradients, may not affect directly the associated particle flow, but instead are due to interparticle interactions interwoven within the complex transition dynamics of state space, allows making the driving forces of one species mutually act on the other, as suggested by the ratchet mechanism. Therefore, the effects like cooperation in the case of parallel gradients or, for anti-parallel gradients, pumping a species against its concentration gradient should become feasible. In the latter case, there would be a negative entropy production for the driven species. The whole entropy production, i.e., that related to the driving and driven species, must of course be positive in accordance with the second law of thermodynamics. The question is: How is this global entropy production by particle flows through the channel related to entropy productions within state space, or in other words, how do sources and sinks of entropy production emerge within the state space, and how do they translate into the entropy production related to particle flows? 
We consider an ensemble of channels. The ensemble averaged whole entropy production consists of that of the channel ensemble and that of the baths the channels are connected with, i.e.,
\begin{equation}
\dot{S}=\dot{S}_{\text{channel}}+\dot{S}_{\text{baths}}\;.
\end{equation}
That of the channel ensemble is expressed by the dynamics of the Shannon entropy: 
\begin{equation}
\dot{S}_{\text{channel}}=\dfrac{d}{dt}\left(-\sum_{\bsigma} P_{\bsigma}(t)\;\ln(P_{\bsigma}(t))\right)\;.
\end{equation}
As we consider the system in the steady state, this component of entropy vanishes, 
\begin{equation}
\dot{S}_{\text{channel}}=0\;.
\end{equation}
As shown above, entropy production in the baths is related to its particle and heat exchange with the system (Equation~(\ref{LocalEntropy})), i.e.:
\begin{eqnarray}
\dot{S}_{\text{baths}}&=&\frac{1}{2}\sum_{\bsigma,\bvarsigma}\dot{S}_{ \bsigma,\bvarsigma} \cr\cr
&=&\frac{1}{2}\sum_{\bsigma,\bvarsigma}-\Delta\epsilon_{\bsigma,\bvarsigma}\; J_{\bsigma,\bvarsigma}\; \;. \label{Schnackenberg}
\end{eqnarray} 
Note that flow and free energy difference concordantly change sign, if state indices are interchanged, which summing up over all demands the factor $1/2$. From now on in the manuscript, we omit the index ``baths'' and superscript $^{(s)}$ as entropy production is always related to that of the baths and dynamics is studied in the steady state. 
It should be remarked that Equation~(\ref{Schnackenberg}) may be written in a more general form, which makes it applicable also to non-steady states. Then, the free energy difference as a driving force is supplemented by the potential difference related to occupation probabilities, which straightforwardly yields in addition to the bath-entropy production the Shannon-entropy production of the system. 
 
To relate the entropy productions within state space to the entropy production by particle flow through the channel, we consider the states that are in the exchange of particles of species $X$ with Bath 2 at the left site of the channel, $\bsigma=(0,\hdots)\rightleftharpoons \bvarsigma=(X,\hdots)$. Jump in rates $\lambda_{\varsigma\leftarrow\sigma}= \lambda_{\varsigma,\bsigma}$
all have the factor $k_+\; c_2^{(X)}$ in common. The corresponding free energy differences may then be rewritten as: 
\begin{eqnarray}
\Delta\epsilon_{\bsigma,\bvarsigma}&=&-\ln(\lambda_{\sigma,\bvarsigma}/\lambda_{\varsigma,\bsigma})\cr \cr
&=&\Delta\epsilon^{(eq)}_{\bsigma,\bvarsigma} -\ln(c_1^{(X)}/c_2^{(X)})\;,\label{FreeenergyDifference}
\end{eqnarray}
with $\Delta\epsilon^{(eq)}_{\bsigma,\bvarsigma}$ as the free energy difference that would be given under equilibrium conditions, i.e., the concentrations in both baths would be $c_1^{(X)}$. 
All other free energy differences, i.e., those not related to particle exchange with Bath 2, show values equivalent to those under equilibrium conditions. Insertion of the free energy differences from Equation (\ref{FreeenergyDifference}) into Equation~(\ref{Schnackenberg}) and applying Equation (\ref{PotentialDiff}) then give:
\begin{eqnarray}
\dot{S}&=&\frac{1}{2}\sum_{\bsigma,\bvarsigma}-\Delta\epsilon^{(eq)}_{\bsigma,\bvarsigma}\; J_{\bsigma,\bvarsigma}+\sum_{X}\ln(c_1^{(X)}/c_2^{(X)}) \underbrace{\sum_{\bsigma, \varsigma| \text{\tiny{exchange $X$ with Bath 2}}} J_{\bsigma,\bvarsigma}}_{=J^{(X)}}\label{EntropyProd1} \\
&=& \frac{1}{2}\underbrace{\sum_{\bsigma,\bvarsigma}-(\phi_{\bsigma}-\phi_{\bvarsigma})\; J_{\bsigma,\bvarsigma}}_{=0}+\sum_{X}\ln(c_1^{(X)}/c_2^{(X)}) J^{(X)}\label{EntropyProd2} \\
&=&\ln(c_1^{(A)}/c_2^{(A)})\;J^{(A)}+\ln(c_1^{(B)}/c_2^{(B)})\;J^{(B)}\; \label{EntropyProd2a}\\
&=&\Delta\mu^{(A)} \;J^{(A)}+\Delta\mu^{(B)}\;J^{(B)}\label{EntropyProd2b}\\
&=&\dot{S}^{(A)}+\dot{S}^{(B)}\label{EntropyProd3}
\end{eqnarray} 
For evaluation of the sum in Equation~(\ref{EntropyProd1}), we used the relation between particle flow and flow in the state space (Equation~(\ref{FlowSpeciesv})). The vanishing sum of entropy productions in Equation~(\ref{EntropyProd2}) for steady state flows in conservative fields (see Equation~(\ref{PotentialDiff})) follows from Kirchhoff's law (Equation~(\ref{Kirchhoff})). Note that $J_{\bsigma,\bvarsigma}=-J_{\bvarsigma,\bsigma}$. 

Equations~(\ref{EntropyProd1}) and (\ref{EntropyProd2a}) demonstrate that the global entropy production in state space is equivalent to the sum of entropy produced by flows of particles through the channel, $\dot{S}^{(X)}=\ln(c_1^{(X)}/c_2^{(X)})\;J^{(X)}=\Delta\mu^{(X)} \;J^{(X)}$, where we introduced the difference of chemical potentials $\Delta\mu^{(X)}=\ln(c_1^{(X)}/c_2^{(X)}) $. Reading it in the other direction, the above equations show that global entropy production by particle flows emerges from entropy production sources in state space. 

\section{Confinement of State Space by Energetic Constraints and their Effect on Two Species Interparticle Interaction}\label{sec3}
 
In the absence of interparticle interactions, the concentration gradients as driving forces could directly affect their associated particle flow, without influence on the other species. In its presence however, these driving forces become integrated into the network of state space transitions. This implies that the driving force of one species may also cross affect transitions of the other species, which is the base of the ratchet mechanism shown in Figure~\ref{Riemann}. However, the ratchet mechanism in our model is restricted to states in which the two species are positioned in a direct neighborhood. This implies for example for the two site channel that only two of nine states are of relevance, $(AB)$ and $BA)$. An option to make the ratchet mechanism work more efficiently is to increase the channel length, as this allows more states with neighboring particles of the two species. Another is to facilitate transitions towards these states by superimposing appropriate energetic constraints on state space. This is achieved by an attractive empty channel ($-E_0<0$) and the avoidance of states in which more particles of one species are present in the channel ($E_X>0$). Note that in case of the three site channel, the latter are differentiated into a long- or 
 short-range interaction. 
These effects are investigated in Figure~\ref{Drivenchannelflow}, in which the coupling of the two species is quantified by the coupling strength $\Delta E=E0=E_A=E_B $, i.e., all energetic constraints are here raised simultaneously. Species $B$ is assigned no concentration gradient, whereas species $A$ has a gradient pointing from Bath 1 to 2. With the increasing gradient of $A$, the flow of $B$ as obtained from Equations~(\ref{FlowSpeciesv}) and (\ref{FlowSpecies}) increases and reaches a maximum, before it decreases. For a vanishing $\Delta E$, a three site channel reveals a moderately higher driving capability of species $A$, when compared to the two site channel, as evident from the flow of $B$. By increasing slightly the coupling strength $\Delta E=E0=E_A=E_B=0.5$, the driving capacity of $A$ enhances for both channel lengths. In this still low coupling range, there is only a moderate superiority of the long- (lr) over the short-range (sr) interaction for the three site channel. 

At a low coupling strength $\Delta E$, flow of the driving species $A$ exceeds by far that of the driven one $B$ (Figure~\ref{Driving_Drivenchannelflow}). With increasing coupling strengths $\Delta E=0\to4\to15$, the flow of the driven species $B$ increases at the 
 cost of that of the driving one $A$. For the two site channel, both flows converge against each other in the strong coupling limit $\Delta E=15$. This also holds for the three site channel and a long-range interaction, 
\begin{equation}
\lim_{\Delta E\to \infty} J^{(A)}=\lim_{\Delta E\to \infty} J^{(B)}
\end{equation}

\begin{figure}[h]
\centering
\includegraphics[width=14cm]{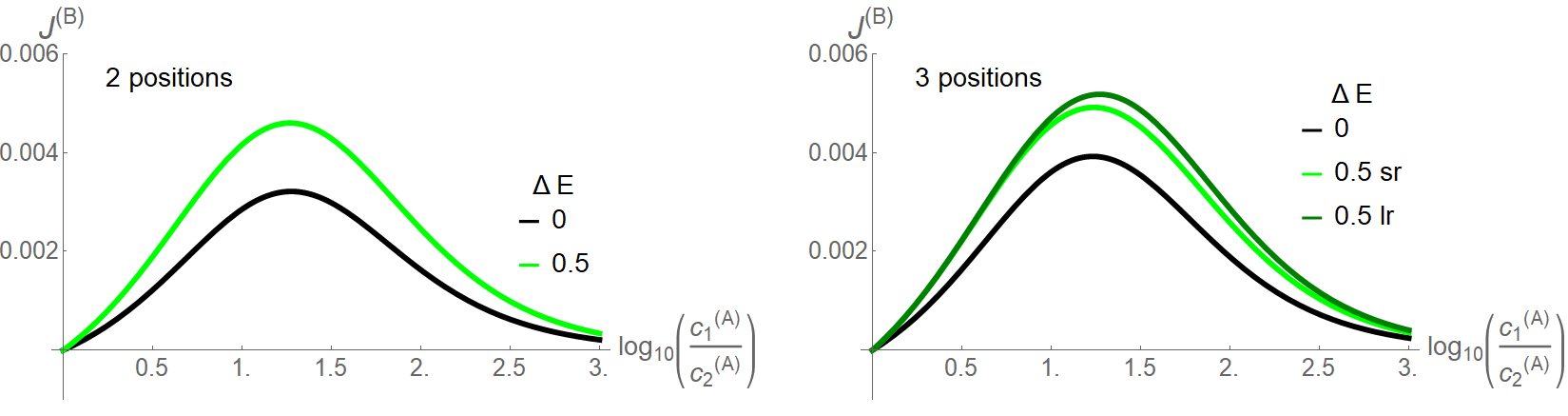}
\caption{Flow of the driven species $B$, $J^{(B)}$, with vanishing concentration gradient ($ k_+ c_2^{(B)}=k_+c_1^{(B)}=0.1,\; k_-=1$ ) as a function of the driving concentration gradient of species $A$, $c_1^{(A)}/c_2^{(A)}$, with $ k_+ c_2^{(A)}=0.1$ held constant. The channel length is =2 positions ({\bf left}) and =3 positions ({\bf right}). Two coupling strengths $\Delta E=E_0=E_A=E_B=0,\; 0.5$ of the two species are shown. Inside the three site channel, a short- (sr) and long-range interparticle interaction (lr) between particles of the same species are differentiated (see text). } 
\label{Drivenchannelflow}
\end{figure}

\begin{figure}[h]
\centering
\includegraphics[width=13cm]{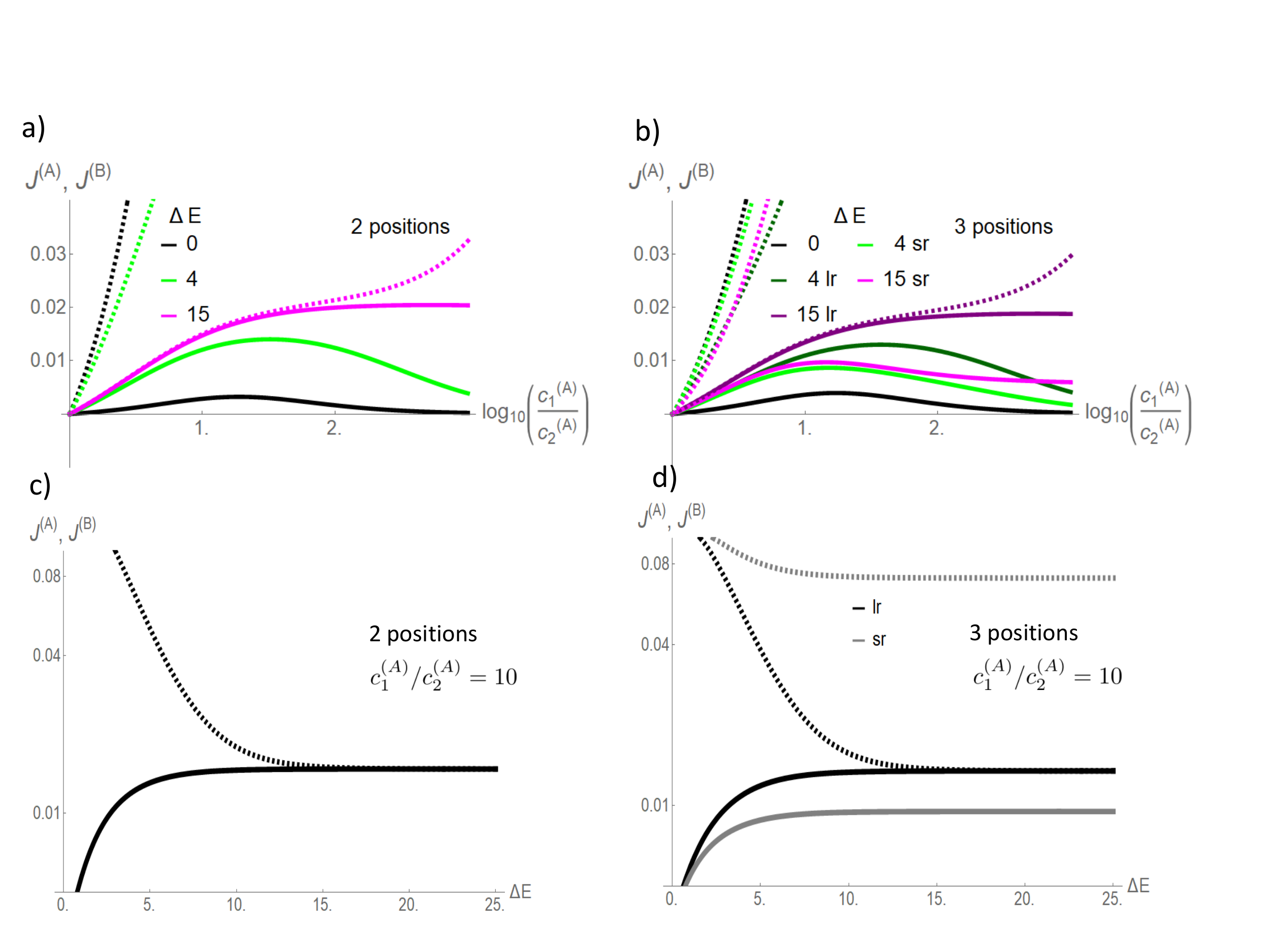}
\caption{Flows of driven ($J^{(B)}$, solid line) and driving species ($J^{(A)}$, dashed line) as a function of the latter's concentration gradient $c_1^{(A)}/c_2^{(A)}$ and exemplary coupling strengths $\Delta E=E_0=E_A=E_B$ for a two site ({\bf a}) and a three site channel ({\bf b}). Other parameters are as in Figure~(\ref{Drivenchannelflow}). The flows of the short-range interactions for the three position channel are in bright, those of the long-range interaction in dark colors. Below: flows (logarithmic scale) are shown as a function of the coupling strength $\Delta E$ for a fixed concentration gradient of the driving species $c_1^{(A)}/c_2^{(A)}=10$ ({\bf c,d}). Dashed curves stand for the driving ($A$), solid for the driven species ($B$); for the three site channel, black lines stand for the long-range interaction (lr), gray lines for the short one. Note the convergence of driving and driven flow with increasing coupling strength for the two site channel and the three site channel with long-range interparticle interaction.}
\label{Driving_Drivenchannelflow}
\end{figure}
\begin{figure}
\centering
\includegraphics[width=16cm]{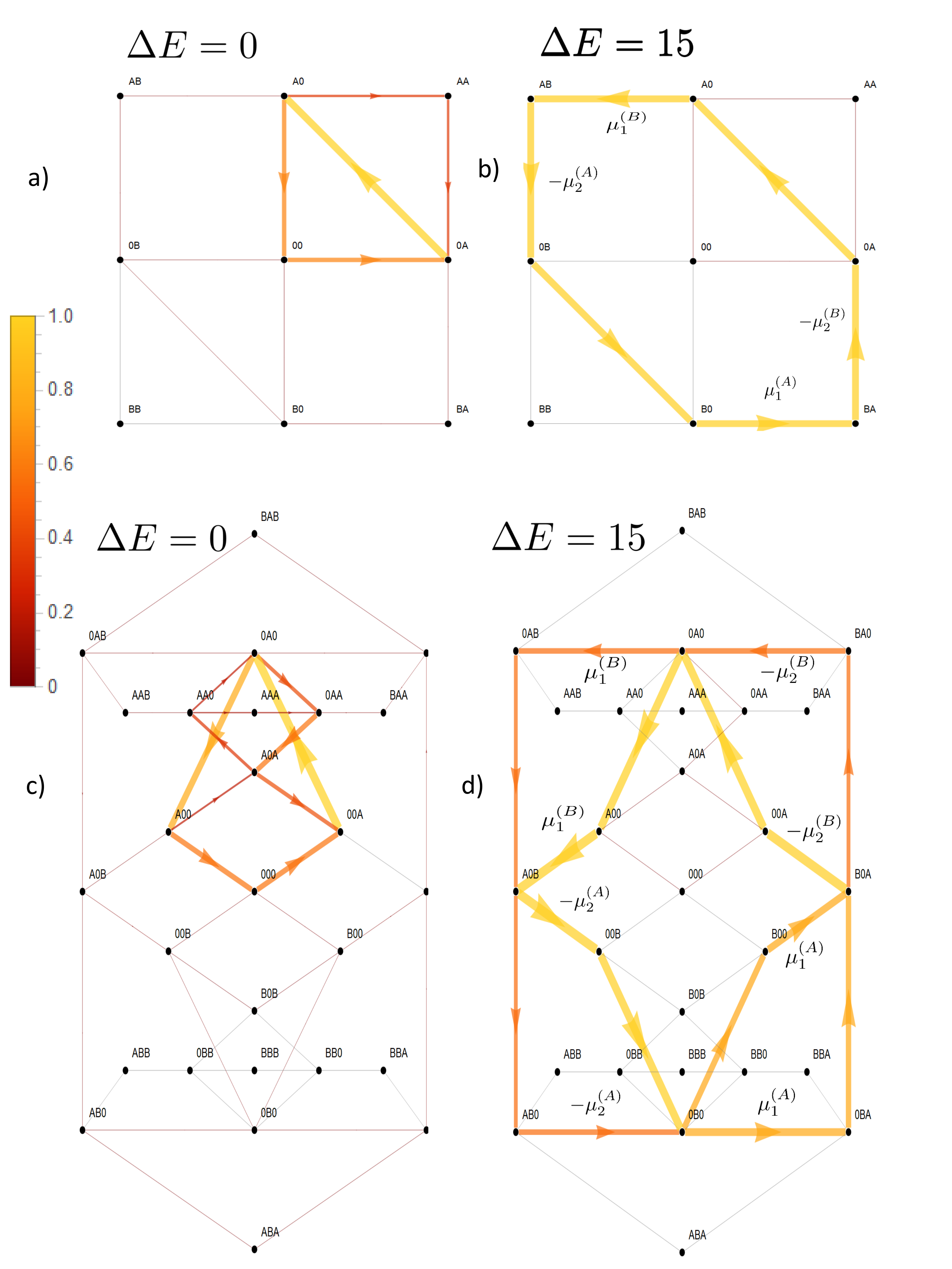}
\caption{Flows in state space for a two site channel ({\bf a, b}) and a three site channel ({\bf c,d}). A vanishing $\Delta E=0$ ({\bf a, c}) and a strong, for the three site channel long-range coupling strength, $\Delta E=15$ ({\bf b, d}) are considered. Flow is normalized to the  maximum magnitude of flow $|J_{max}|$ in state space $ J_{\bsigma,\bvarsigma}\to J_{\bsigma,\bvarsigma}/|J_{max}|$. The magnitude of this normalized flow is coded in colors and in the thickness/size of arrowheads. Flows below $10^{-3}$ are represented by grey lines. For the case of strong coupling, the potentials $\mu^{(X)}_i=\ln(k_+ c^{(X)}_i/k_-)$ are shown next to the transition, which they affect. Concentrations are $k_+ c_1^{(A)}=1,\; k_+ c_2^{(A)}=0.1, \; k_+ c_1^{(B)}=k_+ c_2^{(B)}=0.1$, and the jump off rate is $ k_-=1$.}
\label{StateSpaceFlow_coupling}
\end{figure}

This convergence of flow of the driving and driven species with increasing coupling strength becomes clear in Figure~\ref{StateSpaceFlow_coupling}. In the absence of coupling ($\Delta E=0$), flow is mainly present between states involved in the sole transport of species $A$ (Figure~\ref{StateSpaceFlow_coupling}a,c). For example, for the two site channel, these are the transitions 
\begin{tikzcd}
(A0) \arrow[d,] \arrow[r,]& (AA) \arrow[d,] \\
 (00) \arrow[r,] & (0A)\arrow[lu,] 
\label{DE02P}
\end{tikzcd}
.
Transitions in which species $B$ is involved are negligible. 
An increasing coupling strength elevates the energetic levels of the empty channel state and states occupied by particles of the same species. For large coupling strengths, this hampers visits to these states, which restricts the accessible state space to a subspace with a circular topology (Figure~\ref{StateSpaceFlow_coupling}b,d). In the case of a two site channel, which we will now consider first, a cyclic subspace (CS) emerges (Figure~\ref{StateSpaceFlow_coupling}b). In the steady state, the flow on a cyclic space is constant throughout $\equiv J_{CS}$. In particular, one gets $J_{CS}=J_{(A0),(0A)}=J_{(B0),(0B)} $ and with Equation~(\ref{FlowSpecies}) the equivalence of particle flows :
\begin{equation}
J^{(A)}=J^{(B)}\;|\text{on CS}\;.\label{EquivPFlow}
\end{equation}
As described above, the free energy difference $\Delta\epsilon_{\bsigma,\bvarsigma}$ of a state transition $\bsigma\leftarrow\bvarsigma$ (Equation~(\ref{FreeEnergyLocal})) may be considered as the drift force of this process. On the cyclic subspace, these free energy differences derive from the potentials $ \mu^{(X)}_i=\ln(k_+ c^{(X)}_i/k_-) $ related to particle exchange by $\Delta\epsilon_{(Y,X),(Y0)}=-\mu^{(X)}_1$ and $\Delta\epsilon_{(XY),(0Y)}=-\mu^{(X)}_2$ for the access of $X$ from Bath 1 or 2 to the respective channel end, and with opposite sign, if it leaves. Note that the free energy difference of pure translocations vanishes, $\Delta\epsilon_{(0X),(X0)}=0$. The confinement of state space to the CS makes now the potentials $\mu^{(X)}_i$, and hence, the drift forces act in series 
\begin{tikzcd}
(AB) \arrow[d, "-\mu_2^{(A)}"]& (A0) \arrow[l, "\mu_1^{(B)}",swap] & \\
(0B) \arrow[dr,] & & (0A)\arrow[ul,]\\
& (B0)\arrow[r, "\mu_1^{(A)}"] & (BA) \arrow[u, "-\mu_2^{(B)}",swap]\\
\label{CS2P}
\end{tikzcd}
. Hence, flow on the CS is driven by the free energy difference obtained from the sum of the potentials: 
\begin{equation}
-\mathcal{E}_{\text{CS}}=\mu^{(B)}_1+ ( -\mu^{(A)}_2)+ \mu^{(A)}_1+( -\mu^{(B)}_2)=\Delta\mu^{(B)}+\Delta\mu^{(A)} \label{SumPot}
\end{equation}
This and the equivalence of particle flows in Equation~(\ref{EquivPFlow}) in the case of strong coupling implies that each species is driven by the same force, namely the sum of the chemical potential difference. Hence, the concentration gradient of each species drives to the same amount its own and the other species. 

For a three site channel, the situation is, though a bit more complex, similar as shown in Figure~\ref{StateSpaceFlow_coupling}d. A strong long-range coupling strength $\Delta E $ allows only relevant stochastic transitions between states in which a channel is occupied by a single particle of one or two particles of different species. These states become the elements of the confined state space. Figure~\ref{StateSpaceFlow_coupling}d shows that states of the form $(0Y0)$ and $(Y0X)$ are vertexes of flows, which define a circular graph and, hence, circular topology. Kirchoff's law (Equation~(\ref{Kirchhoff})) implies that flow between these vertexes must be constant in the steady state. Particle flow of, e.g., species $X$ from Bath 1 into the channel, and hence, particle flow through the channel, is equivalent to flow in state space between the vertexes $(0Y0)\dashrightarrow (Y0X) $. The dashed arrow indicates that this flow is the sum of two flows on alternative paths between these vertexes, $ (0Y0)-{{(0YX)-(Y0X)}}$ or $ {{(0Y0-(Y00)}}-(Y0X)$. Both paths differ just by the onset of translocation of $Y$. An interchange of $X$ and $Y$ in the vertexes directly reveals that particle flow of $Y$ from Bath 1 into the channel must be equivalent to that of $X$. Again, we reveal the equivalence of the particle flow of the two species in the limit of strong coupling as in Equation~(\ref{EquivPFlow}).

The free energy difference between the vertexes, which derives from the potentials $ \mu^{(X)}_i=\ln(k_+ c^{(X)}_i/k_-) $, is independent from the the paths between them. Hence, within the circular topology of state space, the potentials act in series 
\begin{tikzcd}
& (0A0)\arrow[dl,dotted, "\mu_1^{(B)}",swap] & \\
(A0B) \arrow[dr,dotted, "-\mu_2^{(A)}"] & & (B0A)\arrow[ul,dotted, "-\mu_2^{(B)}",swap]\\
& (0B0) \arrow[ur,dotted, "\mu_1^{(A)}"]&
\label{CS3P}
\end{tikzcd}
.
Again, the dashed arrows linking these states indicate the two optional paths in between. Hence, as for the two state channel, a strong coupling with a long-range interaction implies that each species is driven by the sum of the chemical potential differences (Equation~(\ref{SumPot})).

However, for the short-range interaction, there is much less driving capacity of $A$ and there is no convergence of particle flows of respective species to each other with increasing coupling strength, as shown in Figure~\ref{Driving_Drivenchannelflow}. This becomes also evident in state space in Figure~\ref{statespaceflow_3P_shortrange}. The short-range interaction leaves the option that the potentials of species $A$ act solely cyclically on its species, which becomes evident on the dominant path in state space $(00A)-(0A0)-(A00)-(A0A)\hdots)$. The last transition would have been impeded in the presence of a long-range interaction. Therefore, only a minor portion of the driving force of $A$ is available for transport of species $B$.

\begin{figure}[h]
\centering
\includegraphics[width=16cm]{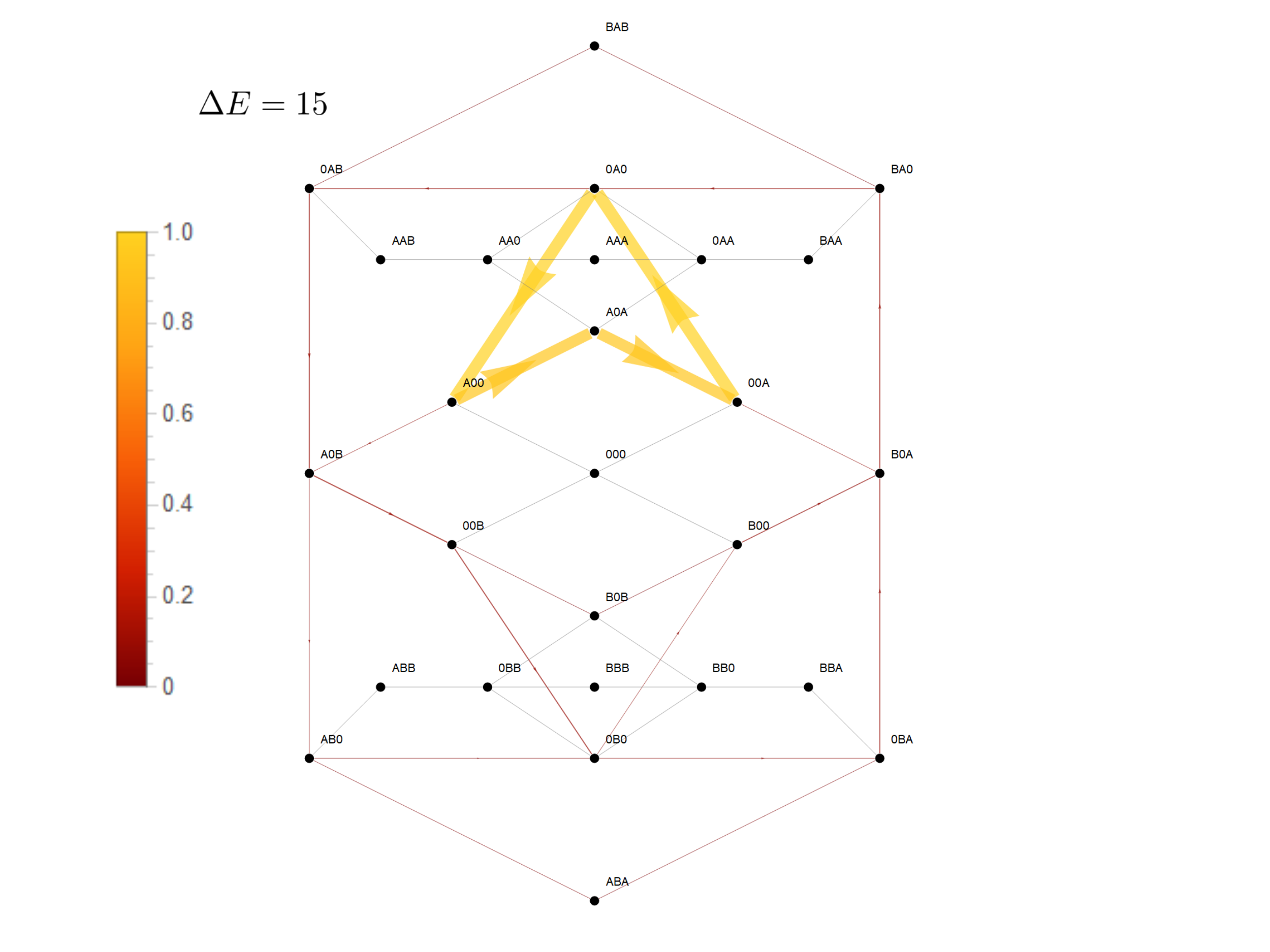}
\caption{Flows in state space for a strong ($\Delta E = 15$) coupling, but only short-range interparticle interaction. Coding of flow, concentrations, and rates are identical to those for the three site channel with long-range interaction in Figure~\ref{StateSpaceFlow_coupling}. }
\label{statespaceflow_3P_shortrange}
\end{figure}


\section{Differential Coupling of the Species and its Effect on Transport }\label{sec4}
In the previous section, a strong coupling of two species implied a strong mutual effect of the driving forces of one species on flow of the other. This was realized by confinement of state space to a subspace with circular topology, in which potentials, and hence driving forces, of the two species are arranged in such a way that they must act in series. For the two site channel, this confined state space is a one-dimensional cyclic space (CS)
\begin{tikzcd}
(AB)\arrow[d, dash]& (A0) \arrow[l, dash] & \\
(0B)\arrow[dr, dash] & & (0A)\arrow[ul, dash]\\
& (B0)\arrow[r, dash] & (BA) \arrow[u, dash]
\label{CS2Pv}
\end{tikzcd}
. To investigate systematically what happens, if transport is less coupled, we will consider an asymmetric situation, leaving transport of species $B$ strongly dependent on that of species $A$, whereas the latter is allowed to bypass the CS. This is realized by a less repulsive interaction of $A$, making visits to the state $(AA)$ more probable. This expands the cyclic state space of strong mutual coupling by a bypass path $-(AA)- $, as seen in Figure~\ref{ReducedStateSpace2P}. This additional path permits species $A$ a leak current on the path $-(AA)-$, which is solely driven by its concentration gradient, with respective free energy reduction $-\Delta\mu^{(A)}$. From the topological point of view, there exist now two entangled cycles: the CS with $\Delta\mu^{(A)}+\Delta\mu^{(B)}$ as the driving force and the cycle 
\begin{tikzcd}
(A0)\arrow[r, dash] & (AA) \arrow[d, dash] \\
& \arrow[lu, dash](0A)
 \label{Leak}
\end{tikzcd}
which makes use of the bypass and on which the system is driven by $\Delta\mu^{(A)}$. Both cycles have the segment $(0A)-(A0) $ in common, the flow on which is identical with particle flow of species $A$ (see Equation~(\ref{FlowSpecies})). In other words: the segment $(0A)-(A0)$ joins two cycles with unequal free energy differences, which determine the flow on this segment. 
\begin{figure}[h]
\centering
\includegraphics[width=6cm]{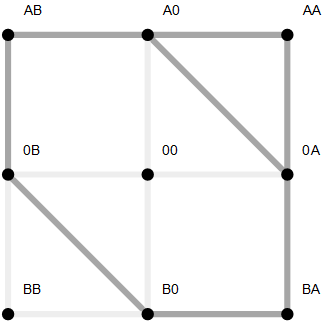}
\caption{Extension of the cyclic state space (CS), to which system transitions are confined in the case of strong coupling ($\Delta E=E_A=E_B=E_0\to \infty$) by the state $(AA)$ (dark gray). This extension is achieved by lowering $E_A$, which enables the occupation of the channel by two particles of species $A$ and by this a leak flow of $A$ on the bypass path $-(AA)-$ in the direction of its concentration gradient. By maintaining a high energy level of the empty channel ($E_0\to\infty$) and strong repulsive interaction between particle of species $B$ ($E_B\to\infty$), flow of $B$ is still confined to the CS. Transitions with negligible flow due to these remaining energetic constraints are shown in light gray. }
\label{ReducedStateSpace2P}
\end{figure}
To see how this differential coupling evolves, we first start with a very strong coupling, $E_A=E_B=E_0=25$ (Figure~\ref{PhaseEntropyA25B25}) in the presence of antiparallel directed concentration gradients, $\Delta\mu^{(A)}>0,\;\Delta\mu^{(B)}<0$. 
\begin{figure}[h]
\centering
\includegraphics[width=16cm]{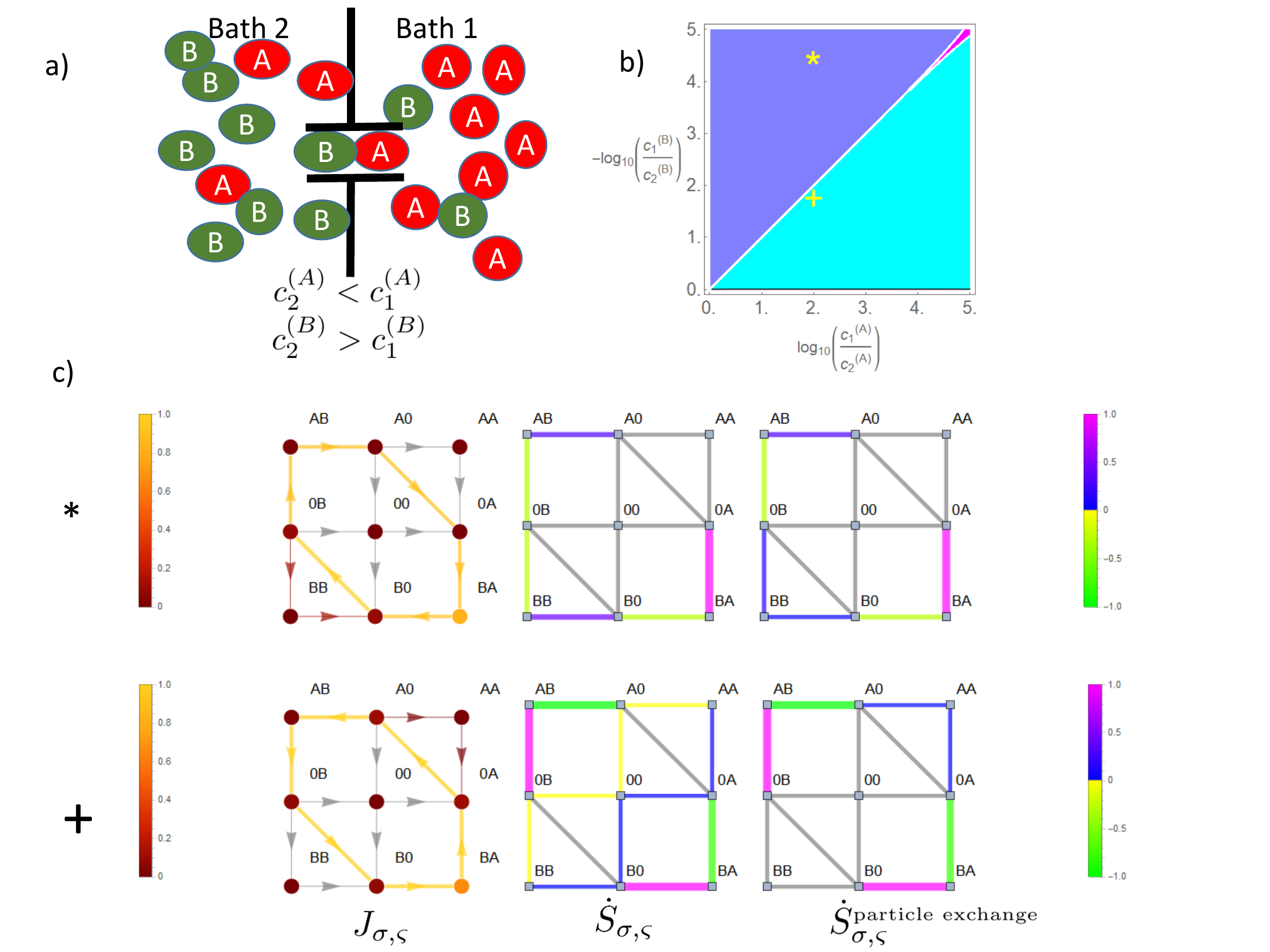}
\caption{Flow dynamics and entropy production in state space under the influence of two opposing concentration gradients. Sketch of gradients and channel in  ({\bf a}). The lower concentration of each species is held constant, $k_+c_2^{(A)}=0.1,\; k_+ c_1^{(B)}=0.1,\; k_-=1$, and the higher concentration $k_+ c_1^{(A)},\; k_+ c_2^{(B)}$ is varied. A strong coupling ($\Delta E=E_A=E_B=E_0=25$) is studied. The phase diagram ({\bf b}) is turquoise if the flow of $A$ is parallel to its gradient and that of $B$ anti-parallel. The opposite holds for the color blue. Magenta stands for gradients in which flows of each species are parallel to its gradient, a phase that is almost absent due to the strong coupling. Two gradient pairs ($ \ast$: $B$ dominating, $+$: $A$ dominating) from this phase diagram are studied in respective rows below ({\bf c}). The left column here shows the solar color coded flow $J_{\bsigma,\bvarsigma}$ and also occupation probability $P_{\bsigma}$ (filled circles) in state space. Note the opposite flow direction on the cyclic space (CS) for the examples $\ast$ and $+$. The column in the middle shows color coded the local entropy production $\dot{S}_{\bsigma,\bvarsigma}=-\Delta\epsilon_{\bsigma,\bvarsigma} J_{\bsigma,\bvarsigma}$ (Equation~(\ref{LocalEntropy})) related to heat and particle exchange, the right column that solely related to particle exchange $\dot{S}_{\bsigma,\bvarsigma}^{\text{particle exchange}}$, i.e., the energy levels $E_X,\;E_0$ for respective transitions are omitted in the free energy difference $\Delta\epsilon_{\bsigma,\bvarsigma} $ (Equation~(\ref{LocalEntropyBathTraj2})). All flows, occupation probabilities, and entropy productions are normalized to its respective maximum magnitude in state space. Values below $10^{-3}$ are shown in gray. }
\label{PhaseEntropyA25B25}
\end{figure}
As shown in the previous section, state space is then almost confined to the CS on which flows of the two species become almost identical $J^{(A)}=J^{(B)}$. The driving forces act in series, i.e., both flows are driven by the sum of chemical potentials $\Delta\mu^{(A)}+\Delta\mu^{(B)}$. This implies 
that in the case of identical magnitude, i.e., $\Delta\mu^{(A)}+\Delta\mu^{(B)}=0$, flow would cease. Otherwise, the flow of both species points in the direction of that with the stronger concentration gradient. This becomes the driving species, which produces positive entropy (Equations~(\ref{EntropyProd2a})--(\ref{EntropyProd3})) by flow through the channel. For the other species, the driven one, the concentration gradient and flow direction are anti-parallel, and hence, entropy production is negative. This dependence of the parallel or anti-parallel orientation of concentration gradient and flow direction, and hence, the sign of entropy production, on the concentration gradient of each species may be best visualized in a phase diagram. For a strong coupling, the phase diagram in Figure~\ref{PhaseEntropyA25B25} shows, besides the curve of vanishing flow at the line of identical magnitude of the gradients, only two phases: the turquoise phase with $J^{(A)}$ parallel and $J^{(B)}$ anti-parallel to its concentration gradient, and for the blue phase, the reverse situation. For each phase, an example with its implications for flow $J_{\bsigma,\bvarsigma} $ and related local entropy production $\dot{S}_{\bsigma,\bvarsigma}=-\Delta \epsilon_{\bsigma,\bvarsigma} J_{\bsigma,\bvarsigma} $ in state space (see Equation~(\ref{Schnackenberg})) is studied: either the gradient of species $B$ ($\ast:\; \Delta\mu^{(B)}+\Delta\mu^{(A)}<0$) or that of $A$ dominates ($+:\; \Delta\mu^{(B)}+\Delta\mu^{(A)}>0$). In Figure~\ref{PhaseEntropyA25B25}, we consider, besides this local entropy production, also that local entropy production solely related to particle exchange with the baths (see Equation~(\ref{LocalEntropyBathTraj2})), i.e., it leaves out potential heat production or absorption related to transitions to and from states at high energy levels, i.e., $E_A, E_B, E_0$. For transitions not including these states, both local entropy productions are identical. Note that there is no entropy production for state transitions related to pure spatial translocations $(X0)\leftrightharpoons (0X)$, as there is no free energy difference. 
Confinement of state space to the cyclic CS implies that flow here is constant throughout, i.e., $J_{CS}\equiv |J_{\bsigma, \bvarsigma |\bsigma, \bvarsigma \in \text{CS} }|$, and in particular, it is equivalent to particle flows (Equation~(\ref{EquivPFlow})). Hence, flow in state space related to bath-channel transitions of the dominating species implies here a local positive entropy production. In the case of species $B$ ($\ast$), this refers to transitions $(0A)\to (BA)$ and $(AB)\to (A0)$. Negative local entropy production is related to bath-channel transitions of the driven species $A$, i.e., $(0B)\to (AB)$ and $(BA)\to (B0)$. The local entropy productions reverse sign if species $A$ becomes the driving one ($+$). Of note is that despite the fact that the high energy barriers confine flow in state space almost to the cyclic state space, there is still some residual flow to and from states of high energy, which explains the small amounts of entropy production for state transitions outside the CS. 

When we reduce the energy barrier of the state occupied by two particles of species $A$, state space expands from the cyclic state space to the reduced state space in Figure~\ref{ReducedStateSpace2P}. This gives rise to a third phase in which a parallel flow and concentration gradient coexist for both species (magenta in Figure~\ref{PhaseEntropyA0B25}), i.e., in this phase, both species produce positive entropy (Equations~(\ref{EntropyProd2a})--(\ref{EntropyProd3})). 
At the dashed lines in the phase diagram in Figure~\ref{PhaseEntropyA0B25}, there is a phase transition between this phase and a phase in which one species is driven against its concentration gradient. Hence, the flow of this latter species ceases here, and phase transition lines are obtained from the equations $J^{(B)}(\Delta\mu^{(A)},\Delta\mu^{(B)})=0$ and $ J^{(A)}(\Delta\mu^{(A)},\Delta\mu^{(B)})=0$, which may be solved analytically as shown in Appendix A. 
There is an important difference between the lines on which the flow of species $B$ vanishes compared to that of $A$. As can be seen from Equations~(\ref{vanishB}) and (\ref{asymptoticB})) and Figure~\ref{PhaseEntropyA0B25} there is an asymptotic gradient of $B$ making its flow cease at high concentration gradients of $A$. The corresponding difference of the chemical potential is: 
\begin{eqnarray}
\Delta\mu^{(B)}_{\infty}&=&\lim_{\Delta\mu^{(A)}\to\infty} \Delta\mu^{(B)}|\text{phase transition}\cr\cr
&=&-\ln\left(1+2\; e^{+E_A/2}\;\frac{1}{k_+ c_2^{(A)}} \right)\;,\label{asymptoticCP}
\end{eqnarray}
which takes for our example ($E_A=0,\; k_+c_2^{(A)}=0.1$ the value $\Delta\mu^{(B)}_{\infty}\approx -3$ or in terms of decimal logarithm $-\log_{10}(c_1^{(B)}/c_2^{(B)})_{\infty}=1.3 $, as shown in Figure~\ref{PhaseEntropyA0B25}. 
Above this gradient $ |\Delta\mu^{(B)}_{\infty}|$, the flow of $B$ cannot be compensated by any gradient of $A$. This is due to the fact that the option of a leak current on the bypass path $-(AA)-$ weakens the driving effect of species $A$, which, in the case of strong coupling, it could otherwise exert on $B$ on the cyclic state space. 

\begin{figure}[h]
\centering
\includegraphics[width=16cm]{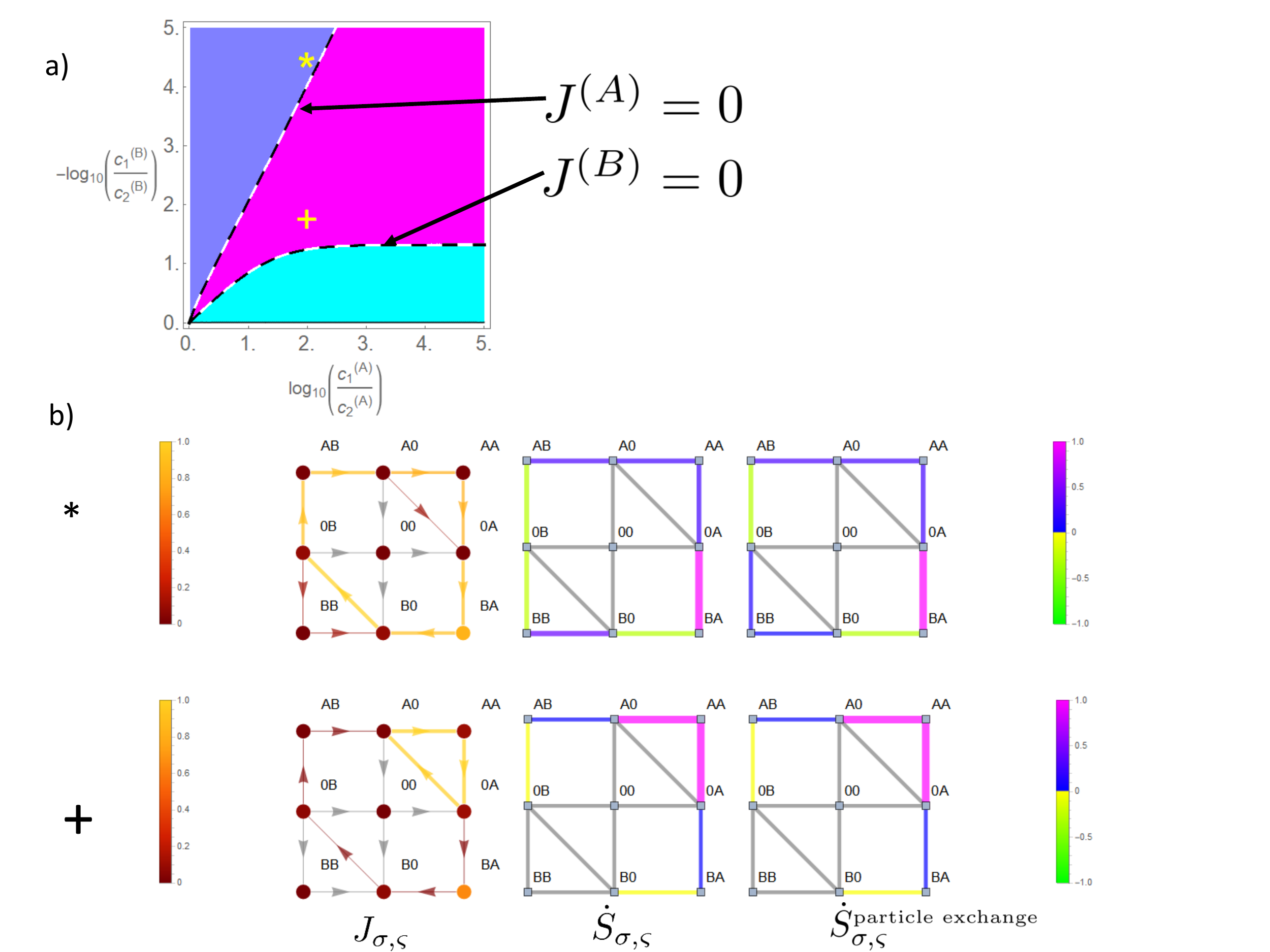}
\caption{Continuation of Figure \ref{PhaseEntropyA25B25}. Transport of species $A$  is less coupled to that of $B$ by setting  $E_A=0$, whereas the other constraints are maintained, $E_B=E_0=25$. In the phase diagram ({\bf a}) a third phase (magenta) emerges in which flow and concentration gradient are parallel for both species. On its boundaries (dashed lines), flow of the species undergoing a change of its flow direction vanishes.  For the two examples $(*)$ and $(+)$ local flow  and entropy production in state space are shown in ({\bf b}). }
\label{PhaseEntropyA0B25}
\end{figure}

In contrast, the flow of $A$ may at any gradient be ceased by an opposing gradient of $B$ as becomes also evident from Equation~( \ref{vanishA2}). The reason is that due to strong coupling flow of species $B$  always implies a parallel directed component of flow of species A. Hence, a sufficient strong gradient of $B$ will cease flow of $A$.

In the example ($\ast$), shown in Figure~\ref{PhaseEntropyA0B25}, species $B$ still maintains its driving capabilities; however, the effect on the flow of $A$ through the channel is reduced, when compared to the situation of strong coupling in Figure~\ref{PhaseEntropyA25B25}. This becomes evident for flow in state space on segment $(A0)\to (0A) $, which is equivalent to the particle flow of $A$ through the channel (Equation~(\ref{FlowSpecies})). This diminished driving effect of $B$ is explained by the option of a leak current $J^{(A)}_{leak}$ of species $A$ in the direction of its concentration gradient on the bypass path 
\begin{tikzcd}
(A0)\arrow[r, "J^{(A)}_{leak}"] & (AA) \arrow[d, "J^{(A)}_{leak}"] \\
& \arrow[lu, "J^{(A)}_{leak}" ](0A)\;
 \label{Leak2}
\end{tikzcd}.
This leak current is directed oppositely to the flow component of $A$, which is driven by species $B$, which results in a diminished magnitude of the net flow of $A$ through the channel.

The sources of entropy production in state space behave accordingly. There is a strong positive entropy production on the bypass path $-(AA)-$ and on transitions in which $B$ moves in the direction of its gradient $(0A)\to (BA)$ and $(AB)\to(A0)$. Negative entropy production appears for transitions of $A$ in state space against its gradient $(0B)\to(AB)$ and $(BA)\to (B0)$. Note: as in the previous example, the high gradient of $B$ allows some residual flow also to and from state $(BB)$ in the direction of the gradient. This generates a positive entropy related to particle exchange and a negative due to heat absorption for the transition $(0B)\to (BB)$.
  
In the example ($+$) in Figure~\ref{PhaseEntropyA0B25}, both species produce positive entropy (magenta colored phase). Hence, species $A$ has lost its driving capabilities, which were present for strong coupling in Figure~\ref{PhaseEntropyA25B25}. Its main flow fraction in state space runs on the bypass path $-(AA)-$, and by this, $A$ loses its impact to drive $B$ against its gradient on the CS. Instead, the flow of $B$ runs parallel to its concentration gradient $(B0)\to (0B) $. 
The leak flow of $A$ on the bypass path produces a large amount of positive entropy. On the cyclic state space, $B$ generates positive entropy on transitions parallel to its gradient and a small amount of negative entropy on transitions driving $A$ against its gradient. 
\begin{figure}[h]
\centering
\includegraphics[width=8cm]{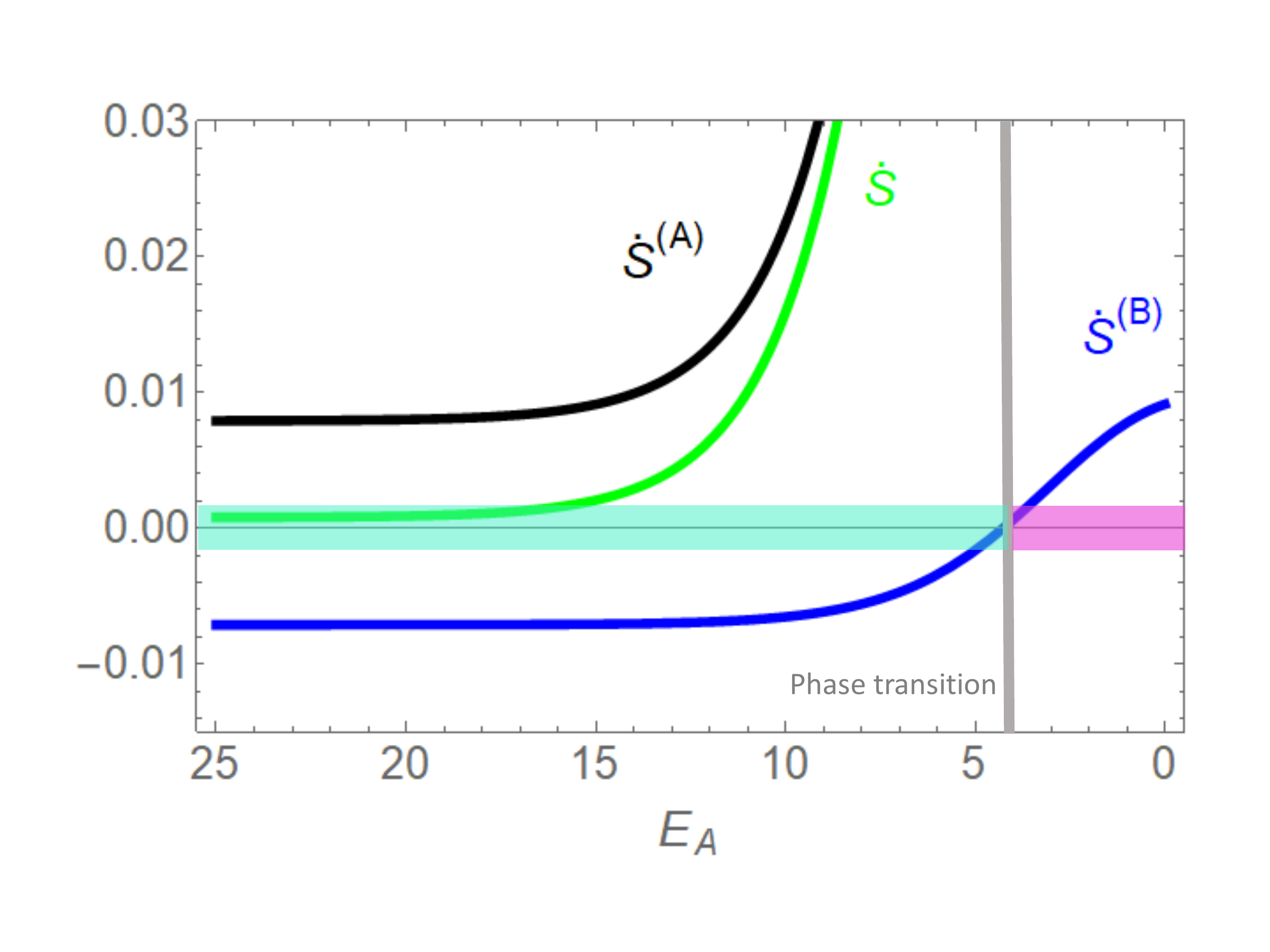}
\caption{Entropy production related to particle flows $\dot{S}^{(A)},\;\dot{S}^{(B)} $ and global entropy production $\dot{S}=\dot{S}^{(A})+\dot{S}^{(B)} $ (see Equations~(\ref{EntropyProd2a})--(\ref{EntropyProd3})) as a function of the option of $A$ to bypass the perfect coupling on the CS. This option is quantified by energy level $E_A$ of the state $(AA)$. The attractive empty channel and the strong repulsive interaction of $B$ are maintained ($E_0=E_B=25$), which leaves transitions in which $B$ is involved confined to the CS. Concentration gradients are that of the example ($+$) in Figures~\ref{PhaseEntropyA25B25} and \ref{PhaseEntropyA0B25}, i.e., $c_1^{(A)}/c_2^{(A)}=10/0.1,\; c_2^{(B)}/c_1^{(B)}=10^{0.8}/0.1$. Entropy production by particle flow is positive, if gradient and flow are parallel, otherwise negative. The higher magnitude of the concentration gradient of $A$ makes it the driving species, with an always positive entropy production. In agreement with the color coding in Figures~\ref{PhaseEntropyA25B25},\ref{PhaseEntropyA0B25}, turquoise stands for the phase of positive entropy production of $A$ and negative entropy production of species $B$. Magenta is the phase in which both species produce positive entropy. The grey line marks the phase transition, $J^{(B)}=0,\; \dot{S}^{(B)}=0$. Global entropy production is always positive according the the second law of thermodynamics.} 
\label{EntropyProdABvsEAA.png}
\end{figure}
In the above examples, it is interesting to see how the flow of $B$, which is equivalent to flow in state space on the remaining CS, is distributed with regard to the leak flow. Kirchhoff's law implies the equivalence of: 
\begin{equation}
\underbrace{J_{(A0),(AB)}}_{=J_{(0B),(B0)}=-J^{(B)}}=\underbrace{J_{(AA),(A0)}}_{=J^{(A)}_{leak}}+\underbrace{J_{(0A),(A0)}}_{=-J^{(A)}}\;,
\end{equation}
i.e.,
\begin{equation}
\underbrace{-J^{(B)}}_{>0}=\underbrace{J^{(A)}_{leak}}_{>0}- J^{(A)}\;.
\end{equation}
Hence, the flow of $B$, being translated into flow in state space, is comprised in the leak flow, either partially in example $(*)$ as $J^{(A)}<0 $ or completely in example $(+)$ as $J^{(A)}>0 $. In between, if phase transition occurs, $J^{(A)}=0$, the magnitude of the flow of $B$ 
 is equivalent to that of the leak flow. Therefore, the bypass offers for $B$ the option that a considerable amount of its transport depends on transitions on the subspace
\begin{tikzcd}
(AB)\arrow[r ]& (A0) \arrow[r] & (AA) \arrow[d] \\
(0B)\arrow[u] & & (0A)\arrow[d]\\
& (B0)\arrow[ul] & (BA) \arrow[l]
\label{CS2Pvv}
\end{tikzcd}
. Summing up the free energy differences of this subspace in the direction of the path shows that after one cycle, the driving free energy difference is 
$\Delta\mathcal{E}=\Delta\mu^{(B)}<0$. Hence, the bypass path enables species $B$ to be solely driven by its gradient on this above subspace. This also explains why species $A$ cannot drive species $B$ against its gradient, if it is above the threshold in Equation~(\ref{asymptoticCP}), which holds for our example $(+)$. Even for high opposing gradients of $A$, the net driving force for species $B$ remains its gradient, i.e., the direction of the flow and gradient remain parallel. 

The sum of entropy produced by the sources in state space (Figures~\ref{PhaseEntropyA25B25} and \ref{PhaseEntropyA0B25}) is equivalent to entropy production by particle flows (Equations~(\ref{EntropyProd2a})-(\ref{EntropyProd3})). In Figure~\ref{EntropyProdABvsEAA.png}, this entropy production is analyzed as a function of $E_A$, i.e., the parameter quantifying the coupling of species $A$ to transport of species $B$, or in geometrical terms: the relation of the leak flow on the bypass path and  flow  on the remaining CS, which is identical with flow of $B$. Concentration gradients are that of the example ($+$) in Figures~\ref{PhaseEntropyA25B25} and \ref{PhaseEntropyA0B25}, $c_1^{(A)}/c_2^{(A)}=10/0.1,\; c_2^{(B)}/c_1^{(B)}=10^{0.8}/0.1$. As already pointed out above, a strong coupling ($E_A=25$) implies (almost) identical particle flows, $J^{(A)}=J^{(B)}=J_{CS} $. The slightly higher magnitude of the gradient of species $A$ makes flow run in its direction, i.e., entropy production related to flow of $A$ ($\dot{S}^{(A)}$) is positive, that of $B$ ($\dot{S}^{(B)}$) negative. Overall entropy production $\dot{S}=\dot{S}^{(A)}+\dot{S}^{(B)}=(\Delta\mu^{(A)}+\Delta\mu^{(B)})J_{CS}$ must be positive in accordance with the second law of thermodynamics. A decreasing $E_{A}$ enables a leak flow of $A$ bypassing the CS on the path $-(AA)-$ in the direction of its gradient and by this a dramatic increase of the respective positive entropy production. However, this leak option attenuates the driving force of $A$. A sufficiently low $E_{A}$ eventually makes the sign of flow of $B$ change in the direction of its gradient, with a phase transition from negative to positive entropy production $\dot{S}^{(B)}$.


\section{Discussion}

Research on the mechanisms of channel transport has the beauty that it covers a broad range of aspects, ranging from very practical descriptive to sophisticated theoretical ones. Whereas the focus of the first is often to provide a detailed model of a real channel, e.g., by simulations, the aim of the latter is to seek for a fundamental understanding of the mechanisms underlying channel transport. Of course, this should not be understood as a dichotomy, as mutual inspiration of both creates a broad spectrum of research in between. 

Coming from the more theoretical view, important factors determining channel transport are particle-channel and inter-particle interactions. There is a huge body of knowledge about how particle-channel interactions affect transport, e.g., by increasing the translocation probability in the case of an attractive force \cite{bezrukov2002, bezrukov2003b, bauer2005, Bauer2006, bauer2010}. Flow of non-self-interacting particles is proportional to this translocation probability, which reveals a permutation symmetry for the location of particle-channel interactions. Hence, flow increases monotonically with binding strength, independent of its localization of the binding site. However, for self-interacting particles, an increasing binding strength leads to blocking of a narrow channel. Therefore, the maximum of flow is reached at a binding strength at which there is a trade-off between both counteracting effects \cite{Bauer2006, Berezhkovskii2005, Berezhkovskii2005b, Bezrukov2007, Kolomeisky}. In the presence of a concentration gradient, blocking depends on the localization of the binding site within the channel, which breaks the symmetry of flow dependence on the location of the binding site. Flow is higher the more the binding shifts in the direction of the gradient \cite{Bauer2006, Bezrukov2007, bezrukov2009}. 

This symmetry breaking effect of the concentration gradient is not only of relevance for blocking, but even more interesting if the self-interaction of particles within the channel becomes feasible. Whereas blocking is just the effect of particle-particle interaction on the access of particles to the channel ends from the baths, the particle-particle interaction within the channel is more subtle. This becomes in particular evident if different species take part in channel transport. For parallel directed concentration gradients, we could recently demonstrate \cite{Bauer2013, Bauer2017} that depending on the magnitude of these gradients, different species may cooperate, i.e., mutually, their flows are higher in the presence of the other one's gradient when compared to flow in its absence. This phase of cooperation is adjacent to phases in which one species promotes the flow of the other at the cost of its own flow, and to a phase at higher gradients, in which mutually, one species hampers the other. We could show in this manuscript that if the gradient of one species vanishes or is even opposing the non-vanishing gradient of the second species, the first experiences a rectifying influence, i.e., it is either driven in the direction of the second or at least its flow is diminished. The mechanism responsible for these mutual rectifying effects is that of a Brownian ratchet. In its original sense \cite{Smoluchowski1912, Feynman1963}, the thought experiment Brownian ratchet should demonstrate the apparent breakdown of the second law of thermodynamics by rectification of motion from the random motion of molecules in a bath. The link between bath and rectified system, the ratchet, is assumed to transform the random motion into a net driving force by an asymmetric potential. The solution of this paradox is that the ratchet itself is subject to thermal motion, which foils the assumed rectification, unless there is a temperature difference between ratchet and bath. 
In our model of channel transport, the asymmetric potential that the rectified species $X$ experiences, for which for simplicity, we assume a vanishing concentration gradient in this discussion here, arises from the concentration gradient of the other species $Y$. The probability to find a particle $Y$ at a position within the channel decreases in the direction of its concentration gradient. As all particles share the type of interparticle interaction that a spatial position is occupied only by one particle, $Y$, if adjacent to $X$, leaves for the latter only the option to move in the opposing direction. For example, for a two site channel with a concentration gradient of $Y$ pointing from the right to the left bath, the probability to find the channel in state $(XY)$ is higher, and by this, the transition $(XY)\to (0Y)$ than that of state $(YX)$ with the associated transition $(YX)\to(Y0)$. Therefore, one is inclined to say that on average, there is an entropic force on $X$ in the direction of the gradient of $Y$, i.e., to the left. However, this naive description blanks out the fact that states $(..Y)$ and $(Y...)$ also hamper access of particles $X$ from the right or left bath, respectively. Therefore, in summary, the effect of $Y$ on $X$ should be balanced, and there should be no net driving force. This is exactly what happens, if one impedes $Y$ to pass the channel and, by this, to produce entropy. Otherwise, the second law of thermodynamics would be violated, and we would have exactly the paradox that the Brownian ratchet at a first glance suggests: rectification of flow without production of entropy. However, how can this formal argument based on the second law of thermodynamics, namely that entropy production by a flow of $Y$ in the direction of its concentration gradient is a prerequisite for the creation of a rectification force on $X$, be understood in terms of the ratchet mechanism? For simplicity, we assume a vanishing concentration of $Y$ in the left bath. An optional sequence of transitions, associated with the flow of $X$ from the right to the left bath and involvement of $Y$, is $(0X)\to (X0)\to (XY)\to (0Y)\to (Y0)\to (YX)\to(0X) $. The asymmetric potential emerges from the above-mentioned gradient related different probabilities of $(..Y)$ and $(Y..) $. However, only the transition $ (YX)\to(0X) $, which in this case is irreversible due to the vanishing concentration of $Y$ in the left bath and which finalizes flow of $Y$ towards the left bath, makes this ratchet potential work and enables the system to start again with the initial state, so that we have the option of a cyclic process driven by flow of $Y$ related entropy production. 

In the above example, the species with the vanishing concentration gradient was the rectified one; the other had the ratchet function. In general, for non-vanishing concentration gradients of both species, mutually, each of them experiences a rectifying force of and acts as a ratchet for the other. We described this complex interaction network by a common state variable of both species and transitions within the framework of a state space. This approach allows correlations between particles of the same and other species, and hence, the respective interparticle interactions become explicit. In mean field approaches, these correlations are neglected, by taking average interparticle interactions, which impedes a closer analysis of stochastic paths and sources of entropy production. The transition dynamics between states depends on their free energy difference. Those stochastic paths in state space are favored, in which free energy 
is reduced, i.e., those with a positive entropy production. This free energy driven course of the paths becomes clear after being projected on the free energy landscape above state space. This energy landscape is similar to the Riemann surface with infinite sheets (Figure~\ref{Riemann}), in which the system is driven successively towards lower free energy levels. The average of these stochastic paths translates into the flow of probability in state space. In this manuscript, we demonstrated that in the steady state, the global entropy production, arising from the concentration gradient driven particle flow through the channel, has its sources in the local entropy productions, determined by the flow of probability between states in state space and the respective free energy difference. In general, the free energy landscape leaves many options for stochastic paths to reduce its free energy. Without any special coupling of the species, paths in state space are favored in which single species transport occurs (Figure~ \ref{StateSpaceFlow_coupling}a,c). The reason is that these paths are shorter, e.g., $(0A)-(A0)-(00)-\hdots $ or $(0A)-(A0)-(AA)-\hdots $ for the two site channel. Therefore, their stochastic flow conductance is higher compared to paths involving the interaction of particles of different species. With respect to the above mutual ratchet mechanism, the question arises how to optimize the free energy difference to have the rectifying forces work most effectively. Intuitively, this is achieved by an optimized coupling of both species and by avoidance of pure single species transport. This was realized by increasing the free energy level of the empty channel and that of channels occupied by several particles of the same species, or in terms of interaction forces an attractive empty channel and repulsive forces between similar neighboring particles. For the longer, three site channel, the latter was differentiated into repulsive forces ranging solely to the nearest neighbor position (short-range) and long-range repulsive forces affecting the whole channel. For the two site channel and the three site channel with long-range interaction, this procedure dramatically confined state space to circular spaces in which the potentials related to the bath concentrations $\mu_i^{(X)}=\ln(k_+\ c_i^{(X)}/k_- )$ are arranged in series, with the effect that the concentration gradient related driving forces $\Delta\mu^{(X)}= \mu_1^{(X)}-\mu_2^{(X)}$ also act in series (Figure~ \ref{StateSpaceFlow_coupling} b,d). This optimum coupling implies that mutually, the driving force of one species also drives the other, i.e., the net driving force for both species is $\Delta\mu^{(A)}+\Delta\mu^{(B)} $. Another consequence is that flows of both species become equivalent. For opposing gradients, which are equal in magnitude, this implies that the flow of both species vanishes. Note that this optimum coupling does not hold for the short-range repulsive interaction in the three site channel. Here, alternative paths of single species transport that bypass the optimum paths of coupling are feasible (Figure~\ref{statespaceflow_3P_shortrange}). 

The fact that in the case of perfect coupling, the flows of both species become identical implies that for opposing gradients, there is always a driving and a driven species, with a parallel flow and gradient direction, and hence positive entropy production for the first and anti-parallel orientation with negative entropy production for the latter. Therefore, there are two phases in the concentration gradient phase diagram, which are separated by a line on which opposing gradients of equal magnitude make flow vanish (Figure~\ref{PhaseEntropyA25B25}).
To study systematically the effect of alternative paths besides those on the cyclic space, to which state space is reduced by perfect coupling, the repulsive interaction between neighboring particles of one species was switched off, whereas that for the other species was maintained. Therefore, the transport of the latter species was still bound to perfect coupling with the first. In contrast, the first had the option to bypass the cyclic space, and entropy could also be produced by a leak current, as shown in Figure~\ref{PhaseEntropyA0B25}. The option of bypassing this cyclic state space of perfect coupling allows as a third scenario. For sufficiently strong gradients, both species may flow in the direction of their concentration gradient, which makes a third, magenta phase emerge in the gradient phase diagram (Figure~\ref{PhaseEntropyA0B25}). On its phase boundaries, the flow of the species undergoing a change in flow direction vanishes. However, there is a decisive difference in the two species. Flow of that species that may bypass the cyclic space may always be ceased by a sufficiently high concentration gradient of the other species. Oppositely, there exist sufficient high gradients of the species perfectly coupled to the CS, for which its flow cannot be ceased by any gradient of the other one. The reason is that the leak flow on the bypass diminishes the rectifying force of this species, which it could otherwise exert on the cyclic space. 

Though our model allowed fundamental insights into the channel transport of two species, many questions remain unsolved. We studied short channels with only two or three sites on which particles may reside. The number of states increases exponentially with the length of the channel, which hampers even numerical treatment. Nevertheless, the basic mechanisms by which mutual rectifying of particle transport is increased become already clear in the two site channel model. The three site channel model even allows introducing a spatial dependent interparticle interaction, with significant consequences, as it was shown that only the long-range repulsive interaction between similar particles allowed an optimal coupling of transport of the two species (Figures~\ref{StateSpaceFlow_coupling} and \ref{statespaceflow_3P_shortrange}). However, our repulsive forces had a very simply spatial dependence. For the three site channel, the short-range interaction abruptly stopped beyond the nearest neighbor, and for the long-range interaction, the force impeded further access of similar particles to an occupied channel independent of the interparticle distance and the number of similar particles, which already resided in the channel. It would be interesting to study more realistic repelling forces especially in longer channels, to answer the question about whether almost perfect coupling is solely dependent on forces that affect the whole channel length, as in our example, or whether there are more sophisticated interactions conceivable. 
Another open field is related to the phase diagrams of gradient and flow direction, and hence the sign of entropy production of the species. These phases in the concentration gradient diagram are separated by lines on which the flow of the species undergoing a change in flow direction vanishes. During phase transition at these lines, flow of this species increased monotonically with its concentration gradient. The question arises about whether for longer channels and more complex interparticle interactions, one might get a scenario in which an increasing gradient reduces flow again after phase transition. This is the characteristics of a Brownian donkey, i.e., a system far from equilibrium in which flow is held at zero and that reacts under the influence of an increasing force (concentration gradient) with a movement (flow) in the opposite direction of the force.

\section{Conclusion}
In this manuscript, we presented a rigorous mathematical treatment of the transport of particles of two species through a narrow channel in terms of stochastic thermodynamics. The model conserved explicitly the spatial correlations of the particles by construction of a state space from the occupation states of the channel and considering the stochastic transitions within. The latter determined the free energy profile from which drift forces derived, which in addition to stochastic forces evolved the system in state space. Within this framework,  sources and sinks  of local entropy production emerged in state space, and we evaluated their relation to particle flows through the channel and its related entropy productions. In particular, we showed how interparticle interactions affected this scenario by constraining state space with consequently differential effects on transport of the species. 

Under non-equilibrium conditions, the interparticle interaction of the two species acted like a Brownian ratchet, i.e., in the direction of its concentration gradient, each species mutually exerted a rectifying force on the other. This mechanism became most efficient by an attractive empty channel and interparticle interactions, which favored a channel occupied by particles of different species, which was realized by a repulsive interaction between particles of the same species. This energetic {{intra}}species constraints result in an {{inter}}species coupling of the transport of the two species. The mapping of the channel's transport dynamics onto state space allows the geometric/topological, or more precisely, as we have a discrete space, the graph derived interpretation of this coupling. In the limiting case of very strong coupling, the accessible state space was confined to a subspace with circular topology. On this subspace, the free energy differences between successive states derive from the potentials $\Delta\epsilon=\mp\mu_i^{(X)}=\mp\ln(k_+ c_i^{(X)}/k_- ) $, which now are arranged in series and induce here a circular steady state flow. The sign depends on whether the flow direction between states implies a particle uptake ($-$) or quitting ($+$ ) of the channel. The free energy decline after one cycle in this subspace is equivalent to the sum of the sign weighted chemical potentials $\Delta \mathcal{E}=-(\Delta\mu^{(A)}+\Delta\mu^{(B)}) $. Therefore, each species is driven by its own concentration gradient and by that of the other. In the steady state, flow is constant throughout on this subspace, and in particular, particle flows of the two species become equivalent. Hence, for opposing concentration gradients, the species with the stronger gradient becomes the driving one, which produces positive entropy; the other species is driven against its concentration gradient and produces negative entropy. If the strong interspecies coupling of transport, i.e., the repulsive intraspecies in-channel interaction, is maintained only for one species ($B$) and loosened for the other ($A$), this enables the latter to flow in the direction of its concentration gradient without being coupled to the transport of the first species, i.e., a leak flow emerges. In geometric/topological terms, the path of this leak flow extends the circular subspace of strong coupling by an additional loop, on which the less coupled species is driven by the the free energy difference $-\Delta\mu^{(A)} $. State space then consists of two joined cycles, which have the segment $(A0)-(0A)$ in common, the flow on which is identical to particle flow $J^{(A)} $ of this species through the channel. Kirchoff's law for steady state flow on this segment $J_{(A0),(0A)}$ implies that this flow is the difference of flows on the two residual cycles, i.e., the difference of leak flow and flow on the remaining original circular subspace, which is identical to the particle flow of the still strongly coupled species $B$, $J^{(B)}$. The option of the leak flow on this bypass path implies a range of concentration gradients, in which both species flow in the direction of its concentration gradient and produce positive entropy. However, the interdependence of leak flow and the flow of the species with the strong coupling $B$ implies that a sufficient high concentration gradient of the latter may eventually always cease the flow of the other one. Conversely, the less coupled species does not have this option.  

\vskip 15pt

Funding: This research was funded by the German Research Foundation (DFG, Grant 396923792) and the Bundesministerium für Bildung und Forschung (BMBF, Grant BMBF01 EO1504. The publication was supported by the Open Access Publication Fund of the University of W\"urzburg. 

\appendix
\section{Derivation of phase transitions lines}
To derive the  lines of transition from the drive/driven -  to  the phase in which flow and respective gradients are parallel, we decompose the reduced state space in Fig.\ref{ReducedStateSpace2P}, which consists of the cyclic state space,  CS: $ (0A)-(A0)-(AB)-(0B)-(B0)-(BA)$ and the path of the leak flow $(0A)-(A0)-(AA)$ into 3 segments. First, the common segment $(0A)-(A0)$, second  that of the residual cyclic state space, rCS : $(AB)-(0B)-(B0)-(BA)$ on which flow of $B$ and partially that of $A$ are present,  and third the bypass segment, bp: $-(AA)-$ on which the leak flow of $A$ occurs. The strategy is to obtain for each linear segment individually its steady state flow and then to combine them according Kirchoff's law. This derives the lines of vanishing particle flows, where state transition happens.

We first study the general case of a linear path with $N$ positions on which stochastic transition dynamics between neighboring positions $i+1\rightleftharpoons i$  are given by rates $r_{i+1,i},\; r_{i,i+1}$.  This path is   supposed to connect two baths each of them serving as source and absorber of particles. Note that for didactic reasons we use the example of particle transitions on the path, which implies that the spatial positions are occupied by a number of particles $p_i$. However the derivation holds in general  for probabilities which is the relevant quantity in our state space. The bath connected to position 1 will be labeled as $0$ and that to positions $N$ as $N+1$, and in accordance the respective transitions rates, and particle concentrations $p_0,\; p_{N+1}$. The transition rates define a free energy difference $\Delta\epsilon_{i+1,i}=-\ln(r_{i+1,i}/r_{i,i+1})$. On a linear path one may assign each position a potential 
\begin{equation}
\varphi_i=\varphi_0+\sum_{\nu=1}^{i}\Delta\epsilon_{\nu,\nu-1}\;, \label{Potentialspath}
\end{equation} 
in which $\varphi_0$ may be set arbitrarily, e.g. zero.      
Flow between neighboring states is obtained from $J_{i+1,i}=r_{i+1,i}p_i-r_{i,i+1} p_{i+1}$. In the steady state flow is constant $J_{i+1,i}\equiv J$. This allows recursively determination of $p_i$ which determines steady state flow as
\begin{equation}
J=\underbrace{(e^{\varphi_0}p_0-e^{\varphi_{N+1}}p_{N+1})}_{\text{difference of activities}} \underbrace{\left(\sum_{\nu=0}^{N}e^{\varphi_{N-\nu}} \frac{1}{r_{N+1-\nu,N-\nu}}\right)^{-1}}_{\text{conductivity}}\;,\label{Ohm}
\end{equation}
which as Ohm's law for diffuse processes implies that flow is proportional to the activity difference and a respective conductivity.  

This is applied to paths in state space.  The bypass path, bp,  and the residual cyclic state space, rCS,  are both adjacent to the positions $(A0)$ and $(0A)$. The probability is stationary in the steady state, i.e. from a formal point of view, these probabilities may be treated like constant concentrations of virtual baths. Note that these ``baths" have nothing to do with the real baths adjacent to the channel ends but are just introduced as a formal mathematical ancillary  construct.  We set  $p_0=P_{A0}$ and $p_{N+1}=P_{0A}$, with $N=1$ for the bypass path and $N= 4$ for the residual positions of the cyclic state space. The respective potentials are obtained from Eq.~(\ref{Potentialspath}), e.g. $\varphi^{(bp)}_{1+1}= -\Delta\mu^{(A)}$ for the bypass path and $\varphi^{(rCS)}_{4+1}=-\Delta\mu^{(A)}-\Delta\mu^{(B)}$ for the residual cyclic state space. Note that for identical start -  $(A0)$ and end point position $(0A)$  the free energy difference of the system depends on the path it has passed. Respective conductivities  are derived from Eq.~(\ref{Ohm}) by insertion of respective transition rates between states in state space  as 
\begin{eqnarray}
C^{(bp)}&=&\frac{1}{2}\;k_{+}c_1^{(A)}  e^{-E_A/2}\cr\cr
C^{(rCS)}&=&\frac{k_-k_+ c_1^{(A)}c_1^{(B)}}{2k_-\;(c_1^{(A)}+c_2^{(A)})+k_-k_+c_1^{(A)}c_2^{(A)}}
\end{eqnarray} 
From Eq.~(\ref{Ohm}) follow the flows in state space 
\begin{eqnarray}
J^{(bp)}&=&\bigg(P^{(s)}_{(A0)}-P^{(s)}_{(0A)}e^{-\Delta\mu^{(A)}}\bigg)\; C^{(bp)}\label{Flowbp}\\
J^{(rCS)}&=&\bigg(P^{(s)}_{(A0)}-P^{(s)}_{(0A)}e^{-\Delta\mu^{(A)}-\Delta\mu^{(B)}}\bigg)\;C^{(rCS)}\label{FlowrCS}
\end{eqnarray}
The flow on the residual CS is equivalent with particle flow of species $B$, i.e. 
\begin{equation}
J^{(B)}=J^{(rCS)}
\end{equation}
For the flow $J_{(0A),(A0)}$ on the common segment $(0A)-(A0)$ which is the particle flow of species $A$  (see Eq.~(\ref{FlowSpecies})) one obtains 
\begin{equation}
J^{(A)}=P^{(s)}_{(0A)}-P^{(s)}_{(A0)}\;.\label{FlowCommonSeg}
\end{equation}
Note that the transition rates $(0X)\rightleftharpoons (X0)$, $\tau_0^{-1}$,  are normalized to 1.
Application of Kirchoff's law states that flow of species $A$ is the sum of the leak flow on the bypass  plus flow on the residual CS (which is that of species $B$), i.e. 
\begin{equation}
J^{(A)}=J^{(bp)}+J^{(B)}\label{KirchoffApp}
\end{equation} 
Combining Eqs.~(\ref{Flowbp},\ref{FlowrCS},\ref{FlowCommonSeg},\ref{KirchoffApp})) determines flows up to a normalization factor $f_n$,
\begin{eqnarray}
J^{(A)}&=&f_n\; \bigg(C^{(bp)}(1-e^{-\Delta\mu^{(A)}})+C^{(rCS)}(1-e^{-\Delta\mu^{(A)}-\Delta\mu^{(B)}})    \bigg)\cr\cr
J^{(B)}&=&f_n\;C^{(rCS)} \bigg((1-e^{-\Delta\mu^{(A)}-\Delta\mu^{(B)}}) +  C^{(bp)}(e^{-\Delta\mu^{(A)}}-e^{-\Delta\mu^{(A)}-\Delta\mu^{(B)}}) \bigg)
\end{eqnarray}
Note that the factor $f_n$ may be obtained from the additional constraint that probability must be conserved in the reduced state space of Fig.\ref{ReducedStateSpace2P}, $\sum_{\bsigma\in\text{reduced state space}} P{\bsigma}=1$, however the tedious derivation is not the scope of this manuscript. 

So we obtain for the phase transition lines for a vanishing flow $B$
\begin{eqnarray}
e^{-\Delta\mu^{(B)}}&=&e^{\Delta\mu^{(A)}}\frac{1+\frac{1}{2}\;e^{\mu_2^{(A)}} e^{-E_A/2}k_-}{1+\frac{1}{2}\;e^{\mu_1^{(A)}} e^{-E_A/2}k_-}\cr\cr
&=&e^{\Delta\mu^{(A)}}\frac{1+\frac{1}{2}\;e^{\mu_2^{(A)}} e^{-E_A/2}k_-}{1+\frac{1}{2}\;e^{\Delta\mu^{(A)}}e^{\mu_2^{(A)}} e^{-E_A/2}k_-}\;,\label{vanishB}
\end{eqnarray}
and for vanishing flow $A$
\begin{eqnarray}
e^{-\Delta\mu^{(B)}}&=& e^{\Delta\mu^{(A)}}\left(1+\frac{1}{2} e^{-E_A/2}\;e^{\mu_2^{(A)}-\mu_1^{(B)}}\;  F(\Delta\mu^{(A)}) \right) \;\text{with} \cr\cr \label{vanishA1}
F(\Delta\mu^{(A)})&=& 4 \sinh(\Delta\mu^{(A)}) + k_-\;e^{\mu_2^{(A)}}(e^{\Delta\mu^{(A)}}-1)\label{vanishA2}
\end{eqnarray}
with $\mu_i^{(X)}=\ln(k_+\;c_i^{(X)}/k_-)$, as the potentials from which the driving forces $\Delta\mu^{(X)}$ derive. 
Some simple consequences of the above phase transition lines are: with  $E_A \to \infty$ which confines state space to the CS, we obtain for both phase transitions lines $-\Delta\mu^{(B)}=\Delta\mu^{(A)}$, i.e. as expected a cessation of flow for opposing gradients which are equivalent in magnitude.  

The maximum gradient of species $B$ which species $A$ can afford to cease is
\begin{eqnarray}
\Delta\mu^{(B)}_{\infty}&=&-\ln\left(\lim\limits_{\Delta\mu^{(A)}\to\infty} e^{\Delta\mu^{(A)}}\frac{1+\frac{1}{2}\;e^{\mu_2^{(A)}} e^{-E_A/2}k_-}{1+\frac{1}{2}\;e^{\Delta\mu^{(A)}}e^{\mu_2^{(A)}} e^{-E_A/2}k_-}\right)\cr\cr
&=&-\ln\left(1+2\;e^{+E_A/2}\;\frac{1}{k_-} e^{-\mu_2^{(A)}} \right)\cr\cr
&=&-\ln\left(1+2\; e^{+E_A/2}\;\frac{1}{k_+ c_2^{(A)}} \right)
\label{asymptoticB}
\end{eqnarray}  
In contrast there is always an opposing gradient of species $B$ which may cease flow of $A$, as the function $F(\Delta\mu^{(A)})$ in Eq.~(\ref{vanishA2}) is monotonous and unbounded with $F(0)=0$, which makes the term in brackets in Eq.~(\ref{vanishA1})  also monotonously and unbounded increase from 1. The more physical reasons are given in the main text.

\end{widetext}

\bibliography{literature}

\end{document}